\documentclass[useAMS,usenatbib]{article}

\usepackage[english]{babel}
\usepackage[utf8]{inputenc}
\usepackage{fancyhdr}
\usepackage{amsmath}
\usepackage{amssymb}
\usepackage{graphicx}
\usepackage{float}
\usepackage[colorinlistoftodos]{todonotes}
\usepackage[margin=0.9in]{geometry}
\usepackage{verbatim}
\usepackage{bm}
\usepackage{caption}
\usepackage{titling}
\usepackage{natbib}
\usepackage{longtable}
\usepackage{threeparttable}
\usepackage{setspace}
\usepackage{enumitem}
\usepackage{subfigure}

\RequirePackage{amsfonts}
\usepackage{physics}
\usepackage{bbm}
\usepackage{booktabs}
\usepackage{amsthm}
\usepackage{mathtools}
\usepackage{url}
\usepackage{hyperref}

\newtheorem{assumption}{Assumption}

\providecommand{\keywords}[1]
{
	\small	
	\textbf{\textit{Keywords---}} #1
}

\newcommand{\BA}{\mathbf{A}}

\newcommand{\BL}{\mathbf{L}}

\newcommand{\Ba}{\mathbf{a}}

\newcommand{\Bl}{\mathbf{l}}

\newcommand{\cANi}{\mathcal{A}(N_i)}

\newcommand{\cAn}{\mathcal{A}(n)}

\newcommand{\cSNi}{\mathcal{S}(N_i)}
\newcommand{\cSN}{\mathcal{S}(N)}

\newcommand{\cSn}{\mathcal{S}(n)}

\newcommand{\Aij}{A_{ij}}

\newcommand{\aij}{a_{ij}}

\newcommand{\Aik}{A_{ik}}

\newcommand{\Ai}{\mathbf{A}_{i}}
\newcommand{\Li}{\mathbf{L}_{i}}

\newcommand{\ai}{\mathbf{a}_{i}}

\newcommand{\overa}{\overline{\Ba}}

\newcommand{\Bbeta}{\boldsymbol{\beta}}
\newcommand{\Bgamma}{\boldsymbol{\gamma}}
\newcommand{\Brho}{\boldsymbol{\rho}}
\newcommand{\Btheta}{\boldsymbol{\theta}}

\newcommand{\bk}{\color{black}}

\newcommand{\indep}{\perp \!\!\! \perp}

\newcommand{\sumjline}{N_i^{-1} \sum_{j = 1}^{N_i}}

\newcommand{\suman}{\sum_{\Ba \in \cAn}}
\newcommand{\sumanline}{\textstyle \sum_{\Ba \in \cAn}}

\title{G-Formula for Observational Studies under Stratified Interference, with Application to Bed Net Use on Malaria}
\author{Kayla W. Kilpatrick$^1$, Chanhwa Lee$^2$, and Michael G. Hudgens$^{2*}$} 
\date{%
    $^1$Department of Biostatistics and Bioinformatics, Duke University, Durham, North Carolina, U.S.A. \\
	$^2$Department of Biostatistics, University of North Carolina at Chapel Hill, Chapel Hill, North Carolina, U.S.A.\\%
	$*$mhudgens@email.unc.edu
}

\begin{document}

\defcitealias{DHS}{MPSMRM, MSP, and ICF International,  2014}

\maketitle
\doublespacing

	\begin{abstract}
		Assessing population-level effects of vaccines and other infectious disease prevention measures is important to the field of public health. In infectious disease studies, one person's treatment may affect another individual's outcome, i.e., there may be interference between units. For example, the use of bed nets to prevent malaria by one individual may have an indirect effect on other individuals living in close proximity. In some settings, individuals may form groups or clusters where interference only occurs within groups, i.e., there is partial interference. Inverse probability weighted estimators have previously been developed for observational studies with partial interference. Unfortunately, these estimators are not well suited for studies with large clusters.
        Therefore, in this paper, the parametric g-formula is extended to allow for partial interference.
        G-formula estimators are proposed for overall effects,  effects when treated, and  effects when untreated. The proposed estimators can accommodate large clusters and do not suffer from the g-null paradox that may occur in the absence of interference. The large sample properties of the proposed estimators are derived assuming no unmeasured confounders and that the partial interference takes a particular form (referred to as `weak stratified interference').
        Simulation studies are presented demonstrating the finite-sample performance of the proposed estimators. The Demographic and Health Survey from the Democratic Republic of the Congo is then analyzed using the proposed g-formula estimators to assess the  effects of bed net use on malaria.
	\end{abstract}
	
	\keywords{Causal inference; G-formula; Herd immunity; Observational studies; Spillover effect}

\section{Introduction}
\label{Intro}

In settings where individuals interact or are connected, 
one individual's treatment status may affect another individual's outcome, 
i.e., interference may be present between individuals \citep{Cox1958}.
Interference is common in infectious disease research. 
For instance, if one individual wears a mask, 
this could affect whether another individual develops COVID-19.
In some settings, it may be reasonable to assume that individuals within a cluster (or group) may interfere with one another, 
but not with individuals in other clusters, 
i.e., there is partial (or clustered) interference \citep{Sobel2006}. 
Clusters might entail households, villages, schools, or other hierarchical structures. 
For instance, when assessing the effect of an intervention or exposure in students, 
it may be reasonable to assume no interference between students in different schools. 
Under this partial interference setting, 
several methods have been proposed for drawing inference about causal estimands of treatment effects; 
e.g., \cite{Tchetgen2012},
\cite{Papadogeorgou2019},
\cite{Barkley2020},
and
\cite{park2022efficient}.

In the presence of interference, 
it is of interest to assess the effect of policies that alter the distribution of treatment in the population. 
For instance, in the Democratic Republic of the Congo, 
public health officials and policymakers may be interested in estimates of malaria risk for different levels of bed net usage in the population.
In observational studies where partial interference is present, 
it may be unlikely that treatment selection among individuals in the same cluster is independent. 
For example, in household studies of vaccine effects, 
we might expect vaccine uptake to be positively correlated between individuals in the same household. 
Therefore, estimands that will be most relevant to policymakers need to account for possible within-cluster treatment selection dependence. 
\cite{Papadogeorgou2019} and \cite{Barkley2020} proposed such estimands and developed corresponding inferential methods using inverse probability weighted (IPW) estimators. 
These IPW estimators entail inverse weighting by an estimated group propensity score. 
Unfortunately, this approach is not well suited for large groups, 
because, in practice, the estimated group propensity score is
often computed by
multiplying individual propensity score estimates across individuals within the same cluster. 
When clusters are large, this product of individual propensity scores
(each of which is between 0 and 1) 
will tend to be very small.
In the absence of interference, a commonly used alternative to the IPW estimator is the parametric g-formula, 
which entails combining outcome regression and standardization \citep{Robins1986, Hernan2006}. 
This paper proposes an extension of the parametric g-formula for observational studies where partial interference may be present, which is better suited for large clusters compared to IPW.

\begin{figure}[ht]
	\centering
	\begin{subfigure}
		\centering
		\includegraphics[scale=0.35]{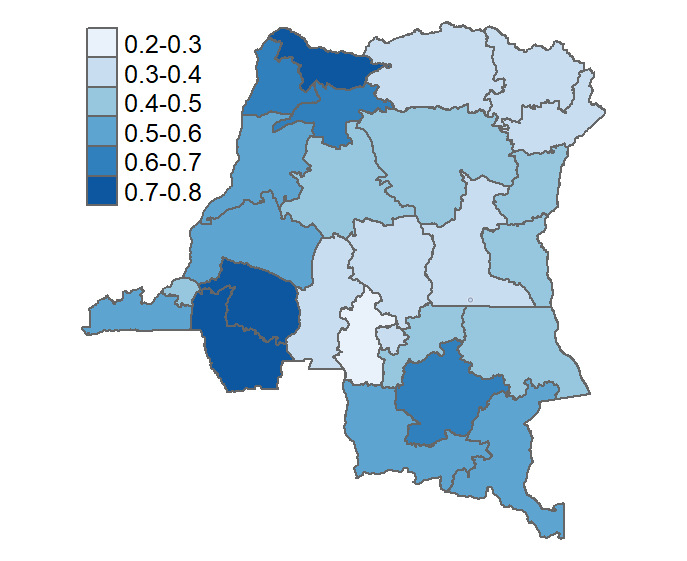}
	\end{subfigure}%
	\begin{subfigure}
		\centering
		\includegraphics[scale=0.35]{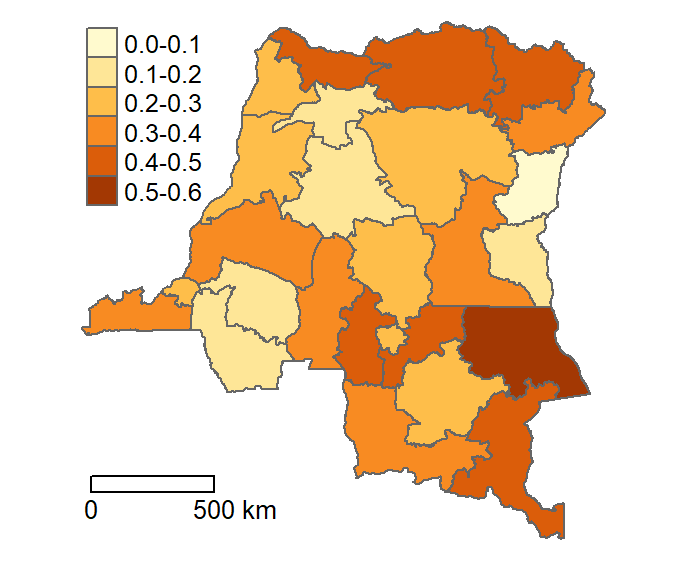}
	\end{subfigure}
	\caption{Malaria bed net study in the Democratic Republic of the Congo. Left map: province-level bed net usage. Right map: prevalence of malaria in children who do not use bed nets.}
	\label{fig:drc heatmaps}
\end{figure}

The proposed methods were motivated by the 2013-14 Democratic Republic of the Congo (DRC) Demographic and Health Survey (DHS), a nationally representative survey to gather information about fertility, maternal and child health, sexually transmitted infections, mosquito net (hereafter ``bed net'') usage, malaria, and other health information \citepalias{DHS}. In the analysis presented below, population-level effects of bed net use on malaria are assessed using data from the DRC DHS. Figure \ref{fig:drc heatmaps} displays province-level bed net use and the proportion of children who did not use bed nets with malaria. The DHS data were collected at the household level. For the analysis here, a single linkage agglomerative cluster method was used to group individuals into clusters based on their household global positioning system (GPS) coordinates, resulting in a total of 395 clusters with at least one child and measured spatial information and other covariates. After performing this clustering algorithm, covariates and bed net use data are available for approximately 87,500 individuals. Malaria outcome data is available for about 7,500 children between 6 to 59 months (for brevity, henceforth referred to as "children"). Among the clusters with at least one child who did not use a bed net, the prevalence of malaria in children who did not use bed nets is inversely associated with the proportion of bed net usage in the cluster (Spearman correlation $r_s=-0.16, p=0.002$), suggesting the possibility of interference within clusters. 
Because malaria is spread between humans via the {\it Anopheles} mosquito, interference is epidemiologically plausible in this setting; 
e.g., if an individual with malaria elects to use a bed net, 
then the likelihood of a mosquito transmitting malaria from that individual to another individual may be decreased.
Previously, 
\cite{Levitz2018} showed that community-level bed net usage was significantly associated with protection against malaria in children younger than five years old. The inferential goal of this paper is to assess the population-level effects of bed use on malaria while allowing for possible within-cluster interference.

The outline of the remainder of this paper is as follows. Section \ref{Methods} presents the proposed extension of the g-formula to allow for partial interference. Section \ref{Sims} presents the simulation results evaluating the performance of the proposed methods in finite samples. In Section \ref{Analysis}, the proposed estimators are employed to assess the effect of bed net use on malaria using data from the DRC DHS. Section \ref{Discussion} concludes with a discussion.

\section{Methods}
\label{Methods}

\subsection{Estimands and Effects of Interest}
\label{estimands} 

Suppose data is observed on $m$ clusters of individuals, and let $N_i$ denote the number of individuals in cluster $i$. 
Suppose some individuals within each cluster may receive treatment (e.g., bed net) and denote the vector of binary treatment indicators in cluster $i$ as 
$\mathbf{A}_i=(A_{i1},A_{i2},\ldots,A_{iN_i})$
with $A_{ij}$ representing the treatment indicator for individual $j$.
Let $S_i=(\sum_{j=1}^{N_i} A_{ij})/N_i$ denote the proportion of treated individuals in cluster $i$. 
Let $Y_i$ represent the outcome at the cluster level. 
In general, $Y_i$ may be defined differently depending on the outcome of interest. 
For example, in the analysis of the DRC data, $Y_i$ may be defined as the proportion of children in a cluster with malaria.
Let $\Li$ represent a vector of cluster-level baseline covariates, including $N_i$. The covariate vector $\Li$ may include summaries of individual-level covariates, such as the average age of individuals within the cluster. 
Let $O_i=\big\{ \Li, S_i, Y_i \big\}$ be the observed random variables for cluster $i$, 
and assume $O_1,\ldots,O_m$ are independent and identically distributed. 
For notational simplicity, the subscript $i$ is omitted when not needed.

Assume partial interference, i.e., there is no interference between clusters, but there may be interference between individuals within the same cluster. 
For example, one individual's bed net usage may affect whether or not another individual in the same cluster gets malaria. 
In the DRC analysis, clusters are defined according to household geographical location.
Thus, the partial interference assumption is biologically plausible as the
{\it Anopheles} mosquito has a limited flight range ($<$ 10 kilometers) and life span ($<$ one month), such that interference (if present) is likely restricted to individuals within the same cluster.
Let $\mathcal{A}(N_i)$ denote the set of all vectors of length $N_i$ with binary entries such that 
$\ai=(a_{i1},a_{i2},\ldots,a_{iN_i})\in \mathcal{A}(N_i)$ 
is a vector of possible treatment statuses for a cluster of size $N_i$. 
For cluster $i$, let $Y_i(\mathbf{a}_i)$ represent the potential outcome if, possibly counter to fact, the cluster had been exposed to $\mathbf{a}_i \in \mathcal{A}(N_i)$, such that $Y_i(\mathbf{a}_i)=Y_i$ when $\mathbf{A}_i=\mathbf{a}_i$.

Population-level effects of interventions such as bed nets can be defined by differences in expected potential outcomes when the distribution of treatment is altered. 
For example, in the absence of interference, the effect of treatment is often defined by the difference in expected outcomes when all individuals receive treatment versus when no individuals receive treatment. 
Here we consider stochastic policies where individuals receive treatment with some probability between 0 and 1. 
Define policy $\alpha$ to be the setting where the expected proportion of individuals in a cluster who receive treatment is $\alpha$, 
i.e., $E_{\alpha}(S) = \alpha$, 
where in general the subscript $\alpha$ denotes the counterfactual scenario in which the policy $\alpha$ is implemented. 
For example, $P_{\alpha} \big( \BA = \Ba | \BL = \Bl \big)$ denotes the conditional probability of treatment given covariates $\Bl$ in the counterfactual scenario in which policy $\alpha$ is implemented.
There are various ways to define the counterfactual treatment allocation probability $P_{\alpha} \big( \BA = \Ba | \BL = \Bl \big)$, 
and the specific policy $\alpha$ considered in this work is described in Section \ref{id}.
% One main distinction from the existing literature on 
Unlike
stochastic policies previously considered in the absence of interference, 
e.g., 
\cite{munoz12}, 
\cite{kennedy19}, 
\cite{wen23},  
the policy considered here allows for within-cluster treatment selection dependence.
For more discussion on other stochastic policies in the presence of partial interference, refer to \cite{lee2023efficient}.
The DRC analysis below considers policies where different proportions of individuals use bed nets. 

Under policy $\alpha$, 
the treatment assignment within a cluster is allowed to be stochastic, 
governed by a distribution on $\BA$ which may vary across clusters depending on cluster-level covariates $\BL$. 
For the stratum of clusters in the population defined by $\BL = \Bl$ with the cluster size $n$, 
the expected potential outcome under policy $\alpha$ is defined by
$
\sumanline
    E \big\{ Y(\Ba) | \BL = \Bl \big\} 
    P_{\alpha} \big( \BA = \Ba | \BL = \Bl \big)
$.
Here, the conditional expectation of the potential outcome $Y(\Ba)$ at stratum of $\BL = \Bl$ is averaged over all $\Ba \in \cAn$,
with corresponding probabilities 
$P_{\alpha} \big( \BA = \Ba | \BL = \Bl \big)$. 
Then, the average potential outcome in the population under policy $\alpha$ is defined as
\begin{align}\label{estimand:mu}
    \mu(\alpha)
    =
    \int_{\Bl}
        \suman
        E \big\{ Y(\Ba) | \BL = \Bl \big\} 
        P_{\alpha} \big( \BA = \Ba | \BL = \Bl \big)
    d F_{\BL}(\Bl)
\end{align}
\bk
where  
$F_{\mathbf{L}}$ denotes the distribution of baseline covariates $\mathbf{L}$. 
Effects of interest can be defined by contrasts in $\mu_{\alpha}$ for two policies $\alpha$ and $\alpha'$, e.g.,
\begin{equation}\label{g formula effect estimand}
\delta(\alpha, \alpha')=\mu_\alpha-\mu_{\alpha'}.
\end{equation}
\bk
\noindent
Here, effects are defined as a difference in average potential outcomes, 
but ratios or other contrasts could be used instead.
A primary contrast of interest in the DRC analysis is the difference in the proportion of children infected with malaria under policies $\alpha$ versus $\alpha'$.

In the DRC analysis, we will consider three different effects of bed nets: the overall effect, the  effect when treated, and the  effect when untreated. All three effects have the form (\ref{g formula effect estimand}) but differ in how $Y_i$ is defined. The overall effect compares the average outcome among all individuals in a cluster under policies $\alpha$ versus $\alpha'$. As it is likely that populations of interest will include a mixture of individuals who would and who would not choose to receive treatment, the overall effect may be valuable for public health officials and policymakers in assessing the overall impact of increasing treatment coverage among a population. 
For inference about the overall effect, $Y_i$ is a summary measure of outcomes in all individuals in cluster $i$. For the malaria data analysis, $Y_i$ is defined to be the proportion of all children in a cluster with malaria.

Two additional treatment effects are also considered. The effect when untreated contrasts average outcomes when an individual is untreated under policy $\alpha$ versus policy $\alpha'$. For this effect, $Y_i$ may be defined by some summary measure of outcomes in untreated individuals. In the DRC analysis of the  effect in the untreated, $Y_i$ will be defined as the proportion of children who do not use bed nets with malaria. If there are no untreated individuals in the cluster, we adopt the convention $Y_i=0$ because no untreated individuals had malaria in those clusters. Similarly, the  effect when treated contrasts average outcomes when an individual is treated under policy $\alpha$ versus policy $\alpha'$. For the  effect when treated in the DRC analysis, $Y_i$ will be the proportion of children who use bed nets with malaria, with $Y_i=0$ in clusters with no treated individuals. 

Causal effects are, in general, defined by contrasts in potential outcomes over the same set of units \citep{Rubin2005}. In this paper, the units are defined to be clusters. While it is natural in many settings to define units as persons or individuals, in the setting considered here, defining units as clusters simplifies matters. In particular, this approach avoids complexities that arise when defining potential outcomes at the individual level, such as nuances regarding casual estimand definitions \citep{VanderWeele2011, Crawford2019}, and drawing inference in a manner that appropriately allows for within-cluster correlation.

Note the effects as defined here could be non-zero due to interference or dependence between individuals' treatment propensity and risk of the outcome. For instance, suppose there is no interference and that individuals with a lower risk of some binary outcome are, in general, less likely to get treatment. Then, as the policy $\alpha$ tends toward 1, only those with the smallest outcome risk will be untreated, such that the mean outcome in the untreated will decrease even in the absence of interference. Nonetheless, the effects defined here could be of interest to policymakers or from a public health perspective, as these effects describe how the average outcome changes across policies.

\subsection{Assumptions and  Identifiability}
\label{id}

The following assumptions are made to identify the estimands described above:

\begin{assumption}[Consistency]\label{assumption:consistency}
    $Y_i(\ai)=Y_i$
    when $\Ai=\ai$.
\end{assumption}

\begin{assumption}[Conditional exchangeability]\label{assumption:exchangeability}
    $Y_i(\ai) \indep \Ai | \Li$
    for all $\ai \in \cANi$.
\end{assumption}

The consistency and conditional exchangeability assumptions are analogous to the assumptions commonly made at the individual level when drawing causal inferences in the absence of interference, 
with the key distinction being that here these assumptions are applied at the cluster level. 
As in the setting where there is no interference, 
selection of the covariates $\mathbf{L}$ such that conditional exchangeability is plausible in a particular application may be informed by subject matter knowledge. 
Causal graphs \citep{Richardson2013}, where the nodes/vertices represent cluster-level random variables, may also be used to determine sufficient sets of covariates $\BL$ for which conditional exchangeability holds.
As discussed in the previous section, the outcome $Y$ may be defined differently depending on the effect of interest.  
Thus, the plausibility of the exchangeability assumption and selection of the particular covariates $\BL$ may differ depending on the choice of $Y$. 
Section \ref{Analysis} discusses the exchangeability assumption in the context of the malaria example.

\begin{assumption}[Weak stratified interference]\label{assumption:wsi}
    $E\big\{
        Y_i(\mathbf{a}_i) | \Li
    \big\}
    = 
    E\big\{
        Y_i(\mathbf{a}'_i) | \Li
    \big\}$
    for all $\ai, \ai' \in \cANi$ such that 
    $\sum_{j=1}^{N_i} a_{ij} = \sum_{j=1}^{N_i} a'_{ij}$.
\end{assumption}

The weak stratified interference (WSI) assumption supposes that the conditional expectation of the cluster-level potential outcome given $\BL$ depends only on the proportion of individuals treated, 
but not which particular individuals receive treatment. 
This assumption is weaker than the usual ``stratified interference'' \citep{Hudgens2008} assumption, 
which stipulates that 
$Y_i(\ai)=Y_i(\ai')$ for any two vectors $\ai,\ai' \in \cANi$ such that $\sum_{j=1}^{N_i} a_{ij} = \sum_{j=1}^{N_i} a'_{ij}$.
The stratified interference assumption might be unrealistic in some settings, 
hence, in this paper, WSI is assumed instead. 
WSI is a weaker assumption in that stratified interference implies WSI, 
but that WSI may also hold in settings where stratified interference does not. 
Note that assuming WSI does not resolve the issue of very small cluster-level propensity scores in large clusters; 
however, under WSI the proposed method permits inference without using inverse probability weighting and thus avoids the extreme cluster-level propensity score issue.

\bk
\begin{assumption}[$S$ model]\label{assumption:Smodel}

    Let
    $\pi
    =
    \pi(\BL;\boldsymbol{\rho})
    =
    g^{-1}(\rho_0 + \boldsymbol{\rho}_1^\top \BL)$,
    where $g$ is some invertible, user-specified link function such as logit or probit,
    and $\boldsymbol{\rho} = (\rho_0,\boldsymbol{\rho}_1^\top)^\top$.
    Assume
    \begin{equation}\label{factual probs}
        P(S = s | \BL)
        =
        P(S =s | \BL; \boldsymbol{\rho})
        =
        {N \choose Ns}
        \pi^{Ns}
        (1-\pi)^{N-Ns}
        .
    \end{equation}
    
\end{assumption}
Note (\ref{factual probs}) will hold if 
$\Aij | \Li$ for $j = 1, \dots, N_i$ are i.i.d. Bernoulli random variables with expectation $\pi(\Li)$, respectively.
In this case,
$\Aij \perp \Aik | \Li$,
i.e., the treatment selection of two individuals within the same cluster is conditionally independent.
This does not, however, imply marginal independence,
i.e., $\Aij \perp \Aik$.
Moreover, provided $\boldsymbol{\rho}_1 \neq 0$, in general we would expect marginal dependence between $\Aij$ and $\Aik$.
While (\ref{factual probs}) is assumed in the rest of this paper, alternative parametric models for the conditional distribution of $S$ given $\BL$ could be assumed instead. 
 
Suppose we are interested in causal estimands corresponding to a counterfactual policy $\alpha$ where the distribution of $S$ is modified such that $\alpha$ proportion of individuals are treated on average. 
In particular, motivated by (\ref{factual probs}),
suppose we would like to draw an inference about the counterfactual scenario where
\begin{equation}\label{counterfactual probs}
    P_\alpha(S = s | \BL)
    =
    P_\alpha(S = s | \BL;\boldsymbol{\gamma})
    =
    \binom{N}{Ns}\pi_{\alpha}^{Ns}(1-\pi_{\alpha})^{N-Ns}
    ,
\end{equation}
where $\pi_{\alpha}
=
g^{-1}(\gamma_{0\alpha} + \boldsymbol{\gamma}_{1\alpha}^\top \BL)$ 
and $\boldsymbol{\gamma}=(\gamma_{0\alpha}, \boldsymbol{\gamma}_{1\alpha}^\top)^\top$
such that 
$E_\alpha(S) = \alpha$. 
The parameter $\Brho$ in (\ref{factual probs}) is identifiable from the observable data, 
whereas the counterfactual parameter $\Bgamma$ in (\ref{counterfactual probs}) are not identifiable without additional assumptions. 
As in \cite{Barkley2020},
assume $\boldsymbol{\rho}_1 = \boldsymbol{\gamma}_{1\alpha}$; 
this assumption implies rank preservation between clusters in treatment propensity. 
In other words, 
if treatment adoption is more likely in cluster $i$ than cluster $j$, 
then under counterfactual policy $\alpha$, 
treatment adoption will also be more likely in cluster $i$ than cluster $j$. 
Then, $\gamma_{0\alpha}$ is obtained from the relationship
$E_\alpha(S) = \alpha$,
by solving the equation
\begin{equation}\label{true gamma 0 alpha}
    \int_{\mathbf{l}}   
        E_\alpha(S|\mathbf{L}=\mathbf{l};\gamma_{0\alpha},\Brho_1)
    dF_{\BL}(\Bl)-\alpha=0
\end{equation}
where
$E_\alpha(S|\mathbf{L}=\mathbf{l};\gamma_{0\alpha},\Brho_1)
=
g^{-1}(\gamma_{0\alpha} + \Brho_{1}^\top \Bl)$. 

\begin{assumption}[$Y$ model]\label{assumption:Ymodel}

    Let
    $\eta=h^{-1}(\beta_0+\Bbeta_1^\top\mathbf{L}+\beta_2S)$ 
    where $h$ is some invertible, 
    user-specified link function,
    and $\Bbeta=(\beta_0, \Bbeta_1^\top, \beta_2)^\top$.
    Assume
    \begin{equation}\label{exp y}
        E(Y|S=s, \mathbf{L}=\mathbf{l})
        =
        E(Y|S=s, \mathbf{L}=\mathbf{l};\Bbeta)
        =
        \eta.
    \end{equation}
    
\end{assumption}
Note assumption \ref{assumption:Ymodel} only supposes that the conditional expectation of $Y$ given $S$ and $\BL$ has a parametric form, with no additional assumptions placed on the conditional distribution of $Y$.  
For simplicity, an interaction between $S$ and $\mathbf{L}$ is omitted from the model of $E(Y|S=s, \mathbf{L}=\mathbf{l})$ but could be included. 
For the analysis in Section \ref{Analysis}, 
binomial regression with the link functions $g=h=\text{logit}$ are used for the treatment and outcome models.

Under assumptions \ref{assumption:consistency} -- \ref{assumption:wsi},
the causal estimands (\ref{estimand:mu}) and (\ref{g formula effect estimand}) are identifiable, i.e., can be expressed as functions of the distribution of the observed random variables.
In particular, from consistency, conditional exchangeability, and WSI, it follows that
$E\{ Y_i(\ai) | \Li \}
=
E \big\{ Y_i | \Ai = \ai, \Li\big\}
=
E \big\{ Y_i | S_i = \overline{\Ba}_i, \Li \big\}$,
where $\overline{\mathbf{a}}_i = \sumjline \aij$.
Therefore, we have the identifiability of $\mu(\alpha)$ as follows:
\begin{align*}
    \mu(\alpha)
    &=
    \int_{\Bl}
        \suman
        E \big\{ Y(\Ba) | \BL = \Bl \big\} 
        P_{\alpha} \big( \BA = \Ba | \BL = \Bl \big)
    d F_{\BL}(\Bl)
    \\
    &=
    \int_{\Bl}
        \sum_{s \in \cSn}
        \sum_{\Ba: \overline{\Ba} = s}
            E \big( Y | S = \overa, \BL = \Bl \big) 
            P_{\alpha} \big( \BA = \Ba | \BL = \Bl \big)
    d F_{\BL}(\Bl)
    \\
    &=
    \int_{\Bl}
        \sum_{s \in \cSn}
            E \big( Y | S = s, \BL = \Bl \big) 
            P_{\alpha} \big( S = s | \BL = \Bl \big)
    d F_{\BL}(\Bl)
\end{align*}
where 
$\cSn = \{0/n, 1/n,\dots, n/n\}$,
and
$E \big( Y | S = s, \BL = \Bl \big) $
and
$F_{\BL}(\Bl)$
are identifiable from the observed data
$O_i=\big\{ \Li, S_i, Y_i \big\}, i = 1, \ldots, m$,
and $P_{\alpha} \big( S = s | \BL = \Bl \big)$ is also identifiable since the parameters $\Brho$, $\gamma_{0\alpha}$, and thus
$\pi_{\alpha}
=
g^{-1}(\gamma_{0\alpha} + \boldsymbol{\rho}_{1}^\top \BL)$ 
are identifiable.
The identifiability of the causal effect
$\delta(\alpha, \alpha')$
can be shown similarly.
Note assumptions \ref{assumption:Smodel} -- \ref{assumption:Ymodel} are not required for identifiability, but are assumed to facilitate inference as described in the next section. 
\bk

\subsection{Inference}\label{estimators}
Estimators for $\mu(\alpha)$ can be constructed as follows. 
First estimate the parameters $\Brho=(\rho_0,\Brho_1^\top)^\top$ of model (\ref{factual probs}) 
and $\Bbeta=(\beta_0,\Bbeta_1^\top,\beta_2)^\top$ of model (\ref{exp y}) via maximum likelihood; 
denote these estimators by 
$\hat{\Brho}=(\hat{\rho}_0, \hat{\Brho}_1^\top)^\top$ 
and $\hat{\Bbeta} = (\hat{\beta}_0, \hat{\Bbeta}_1^\top, \hat{\beta}_2)^\top$. 
Next, for a given policy $\alpha$, 
let $\hat{\gamma}_{0\alpha}$ denote the estimator of $\gamma_{0\alpha}$ obtained by finding the solution to (\ref{true gamma 0 alpha}) with $F_{\mathbf{L}}$ replaced by its empirical distribution, 
i.e., 
$m^{-1}\sum_{i=1}^m     
    \hat{E}_\alpha(S_i|\mathbf{L}_i;\gamma_{0\alpha},\hat{\Brho}_1)
- \alpha
= 0$ 
where 
$\hat{E}_\alpha(S_i | \Li;\gamma_{0\alpha},\hat{\Brho}_1) 
=
g^{-1}(\gamma_{0\alpha} + \hat{\Brho}_1^\top \Li)$. 
Let 
$\hat{P}_\alpha(S=s|\mathbf{L})$ 
denote (\ref{counterfactual probs}) evaluated using 
$(\hat{\gamma}_{0\alpha}, \hat{\Brho}_1^\top)^\top$, 
and let 
$\hat{E}(Y|S=s, \mathbf{L}=\mathbf{l})$
denote (\ref{exp y}) evaluated using $\hat{\Bbeta}$. 
Then the g-formula estimator of $\mu(\alpha)$ is 
\begin{equation*}
    \hat{\mu}(\alpha)
    =
    \int_{\mathbf{l}}
        \sum_{s \in \cSn} 
        \hat{E}(Y|S=s,\mathbf{L}=\mathbf{l})
        \hat{P}_\alpha(S=s|\mathbf{L}=\mathbf{l})
    d\hat{F}_\mathbf{L}(\mathbf{l})
\end{equation*}
where $\hat{F}_\mathbf{L}$ denotes the empirical distribution function of $\mathbf{L}$. Equivalently, the estimator may be written
\begin{equation*}
    \hat{\mu}(\alpha)
    =
    \frac{1}{m}
    \sum_{i=1}^{m} 
        \sum_{s \in \cSNi} 
            \hat{E}(Y_i | S_i = s, \Li)
            \hat{P}_\alpha(S_i = s | \Li)
    .
\end{equation*}	
The estimator for the effects of interest is 
$\hat{\delta}(\alpha, \alpha')
=
\hat{\mu}(\alpha)
-
\hat{\mu}(\alpha')$. 
The estimators 
$\hat{\Brho}, \hat{\Bbeta}$, 
$\hat{\gamma}_{0\alpha}$,
$\hat{\mu}(\alpha)$, $\hat{\mu}(\alpha')$, 
and $\hat{\delta}(\alpha, \alpha')$ 
are solutions to unbiased estimating equations (see the  Appendix). Therefore, it follows from standard large-sample estimating equation theory that the estimators are consistent and asymptotically Normal \citep{Stefanski2002}. The empirical sandwich estimators, which are consistent estimators of the asymptotic variances, can be used to construct point-wise Wald confidence intervals (CIs).

\subsection{Population Strata}\label{subgroups}
For the DRC malaria example, 
the methods described above may be applied directly if children are considered the population of interest, and we ignore data collected from adults. 
Such an approach makes inference about counterfactual scenarios regarding the distribution of bed net usage in children and is agnostic to bed net use by others in the clusters. 
However, the DRC DHS includes bed net data for all individuals,
which can be utilized to estimate the effects of bed net usage by all individuals on the risk of malaria in children. 
To do so, the approach above can simply be modified by changing the definition of $S$ to be the proportion of all individuals in the cluster, not just children, who use bed nets. 
Alternatively, one may choose to model separately the proportion of children using bed nets (say $S_1$) and the proportion of other individuals in the cluster using bed nets (say $S_2$). 
In particular, the population mean estimand $\mu(\alpha)$ may be expressed 
\begin{equation*}
    \int_\mathbf{l} 
        \sum_{s_2 \in \mathcal{S}_2}
        \sum_{s_1 \in \mathcal{S}_1} 
            E(Y|S_1 = s_1, S_2 = s_2, \mathbf{L}=\mathbf{l})
            P_\alpha(S_1=s_1|\mathbf{L}=\mathbf{l},S_2=s_2)
            P_\alpha(S_2=s_2|\mathbf{L}=\mathbf{l})
    dF_{\mathbf{L}}(\mathbf{l})
\end{equation*} 
where the policy $\alpha$ here is defined such that individuals in strata $1$ and $2$ are treated with the same probability: $E_\alpha(S_1)=E_\alpha(S_2)=E_\alpha(S)=\alpha$. More generally, one could consider different policies $\alpha_1$ and $\alpha_2$ for the two population strata. Inference proceeds analogous to Sections \ref{id}--\ref{estimators}, but with separate parametric models for $S_1$ given $\mathbf{L},S_2$ and for $S_2$ given $\mathbf{L}$; such an approach is taken in the DRC bed net analysis in Section \ref{Analysis}. 

\subsection{G-Null Paradox}\label{g-null}
In the absence of interference, the parametric g-formula may give rise to the so-called g-null paradox. That is, certain parametric models are guaranteed to be misspecified under the null hypothesis of no treatment effect. As a result, the null hypothesis of no treatment effect will be incorrectly rejected with high probability when the sample size is large \citep{Robins1986, Robins1997proceedings, Taubman2009}.

For the setting considered in this paper, 
the null hypothesis is that the proportion treated $S$ has no effect on the outcome $Y$, 
or that $\mu(\alpha)=\mu(\alpha')$ for any two policies $\alpha, \alpha'$. 
If $S$ has no effect on $Y$, 
then $\beta_2=0$ and $E(Y|S=s, \mathbf{L})=E(Y|\mathbf{L})$. 
Therefore,
\begin{equation}\label{g null equation 1}
    \mu(\alpha)
    =
    \int_\mathbf{l} 
        E(Y|\mathbf{L}=\mathbf{l})
        \sum_{s \in \cSn} 
            P_\alpha(S=s|\mathbf{L}=\mathbf{l})
    dF_\mathbf{L}(\mathbf{l})
    =
    \int_\mathbf{l} 
        E(Y|\mathbf{L}=\mathbf{l})
    d F_\mathbf{L}(\mathbf{l})
\end{equation}
where the second equality follows because $\sum_{s \in \cSn} P_\alpha(S=s|\mathbf{L}=\mathbf{l})=1$. The right-hand side of (\ref{g null equation 1}) does not depend on $\alpha$, so the g-null paradox does not occur here.

\section{Empirical Evaluation} \label{Sims}
Simulation studies were conducted to evaluate the finite sample properties of the proposed g-formula estimator. Three separate simulation studies were conducted for the three target estimands: overall effect,  effect when treated, and  effect when not treated. For the overall effect simulation study, 1000 data sets each with $m=125$ clusters were stochastically generated as follows:
\begin{enumerate}

    \item[(i)] The number of individuals per cluster $N_i$ was simulated such that $P(N_i=8)=0.4, P(N_i=16)=0.35, \text{ and } P(N_i=20)=0.25$.
    
    \item[(ii)] Two cluster-level covariates $L_{1i}$ and $L_{2i}$ were generated, where $L_{1i}$ was Normal with mean 40 and standard deviation 10, and $L_{2i}$ was such that $P(L_{2i}=0)=5/18, P(L_{2i}=1)= 3/18, P(L_{2i}=2)=4/18, P(L_{2i}=3)=5/18, P(L_{2i}=4)=1/18$. 
    
    \item[(iii)] For each cluster, the number of treated individuals was drawn from a Binomial distribution with parameters $N_i$ and $\pi_i=\text{expit}(\rho_0+\rho_1L_{1i}+\rho_2L_{2i})$ where $\Brho=(\text{logit}(0.6), -0.01, -0.01)^\top$. 
    The proportion of individuals treated per cluster, $S_i$, was then calculated by dividing the number of treated individuals by $N_i$. 
    
    \item[(iv)] For each cluster, the outcome $Y_i$ was set equal to $X_i/N_i$ where $X_i$ was Binomial with parameters $N_i$ and $\eta_i=\text{expit}(\beta_0+\beta_1L_{1i}+\beta_2S_i+\beta_3L_{2i})$ 
    where $\Bbeta=(\text{logit}(0.6), -0.01, -0.8, -0.01)^\top$. 
    
\end{enumerate}

Correctly specified models of $Y$ given $S$ and $\mathbf{L}$, and of $S$ given $\mathbf{L}$ were fit by maximum likelihood. The asymptotic variance of the estimators was estimated using the empirical sandwich variance estimator, and point-wise Wald 95\% CIs were calculated with these variance estimates.

The true values of estimands for policies $\alpha \in \{0.4, 0.5, 0.6\}$ were calculated analytically for the data generating process described above. 
In particular, the true values of $\gamma_{0\alpha}$ are the solutions to (\ref{true gamma 0 alpha}) where $\pi_\alpha=\text{expit}(\gamma_{0\alpha}+\rho_1 L_1+\rho_2L_2)$.
The counterfactual probabilities $P_\alpha(S=s|\mathbf{L})$ for $s \in \cSN$ can then be computed via (\ref{counterfactual probs}) based on the true values of $\gamma_{0\alpha}, \rho_1, \rho_2$.  
Similarly, $E(Y|S=s,\mathbf{L})$ for $s \in \cSN$ may be evaluated using (\ref{exp y}) and the true value of $\Bbeta$.
Finally, the true values of $\mu(\alpha)$ can be found using (\ref{estimand:mu}).

Results for the overall effect simulation study are given in the top third of Table \ref{sim table}. 
The average bias of the proposed g-formula estimators was negligible, and the CIs contained the true parameter values for approximately 95\% of the simulated datasets. 
The average of the estimated sandwich standard errors was approximately equal to the empirical standard errors, with standard error ratios of approximately 1.

The simulation study described above was repeated for the  effect when treated, with the following modification. 
In step (iv), the cluster outcome $Y_i$ was set equal to $X_i^1/(N_iS_i)$ 
where $X_i^1$ was Binomial with parameters $N_iS_i$ and $\eta_i$. 
If there were no treated  individuals in a cluster, then $Y_i$ was set to 0. 
Results for the g-formula estimator of the effect when treated are presented in the middle part of Table \ref{sim table}. 
Results are similar to the overall effect, except the standard error for the g-formula estimator of the  effect when treated is larger because fewer individuals contribute to the outcome.

Finally, a third simulation study was conducted for the  effect when untreated. 
The simulation steps above were repeated, but with step (iv) modified such that the cluster outcome $Y_i$ was set equal to $X_i^0/\{N_i(1-S_i)\}$ where $X_i^0$ was Binomial with parameters $N_i(1-S_i)$ and $\eta_i$, 
with $Y_i$ set to 0 if $S_i=1$. 
Results are given in the bottom section of Table \ref{sim table}.

\begin{center}
	\begin{table}[ht]
		\centering
		\caption{Summary of simulation study results as described in Section \ref{Sims}. Truth: true value of the estimand targeted by the estimator. Bias: average bias of the g-formula estimates over 1000 datasets. Cov\%: empirical coverage of Wald 95\% CIs. ASE: average of estimated sandwich standard errors. ESE: empirical standard error. SER: ASE/ESE.}
		\label{sim table}
		\begin{tabular}{c c c c c c c}
			\hline
			Estimand & Truth & Bias & Cov\% & ASE & ESE & SER\\
			\hline
			\multicolumn{7}{c}{All Individuals}\\
			\hline
			$\mu(\alpha=0.4)$ &  0.418 & 0.000 & 94\%& 0.0147 & 0.0153 & 0.96 \\
			$\mu(\alpha=0.5)$ & 0.399 & -0.000 & 94\% & 0.0119 & 0.0121 & 0.98\\
			$\mu(\alpha=0.6)$ & 0.380 & -0.000 & 94\% & 0.0145 & 0.0149 & 0.97 \\
			$\delta(\alpha=0.6,\alpha'=0.4)$ & -0.038 & -0.001 & 94\% & 0.0172 & 0.0180 & 0.95\\
			$\delta(\alpha=0.6,\alpha'=0.5)$ & -0.019 & -0.000 &94\% & 0.0084& 0.0089&0.95\\
			$\delta(\alpha=0.5,\alpha'=0.4)$ & -0.019 & -0.000 & 94\%& 0.0087& 0.0091&0.96\\
			\hline
			\multicolumn{7}{c}{When Treated}\\
			\hline
			$\mu_1(\alpha=0.4)$ &  0.418 & -0.002 & 95\%& 0.0243 & 0.0242 & 1.00 \\
			$\mu_1(\alpha=0.5)$ & 0.399 & -0.001 & 96\% & 0.0174 & 0.0165 & 1.05\\
			$\mu_1(\alpha=0.6)$ & 0.380 & 0.000 & 95\% & 0.0184 & 0.0178 & 1.03 \\
			$\delta_1(\alpha=0.6,\alpha'=0.4)$ & -0.038 & 0.002 & 93\% & 0.0255 & 0.0267 & 0.96\\
			$\delta_1(\alpha=0.6,\alpha'=0.5)$ & -0.019 & 0.001 &93\% & 0.0126& 0.0132&0.96\\
			$\delta_1(\alpha=0.5,\alpha'=0.4)$ & -0.019 & 0.001 & 93\%& 0.0129& 0.0135&0.96\\
			\hline
			\multicolumn{7}{c}{When Untreated}\\
			\hline
			$\mu_0(\alpha=0.4)$ &  0.418 & -0.001 & 95\%& 0.0185 & 0.0188 & 0.99 \\
			$\mu_0(\alpha=0.5)$ & 0.399 & -0.000 & 96\% & 0.0173 & 0.0167 & 1.03\\
			$\mu_0(\alpha=0.6)$ & 0.380 & 0.000 & 96\% & 0.0235 & 0.0231 & 1.02 \\
			$\delta_0(\alpha=0.6,\alpha'=0.4)$ & -0.038 & 0.001 & 94\% & 0.0248 & 0.0259 & 0.96\\
			$\delta_0(\alpha=0.6,\alpha'=0.5)$ & -0.019 & 0.000 &94\% & 0.0122& 0.0127&0.96\\
			$\delta_0(\alpha=0.5,\alpha'=0.4)$ & -0.019 & 0.000 & 94\%& 0.0126& 0.0131&0.96\\
			\hline
		\end{tabular}
	\end{table}
\end{center}

Additional simulation studies were conducted to compare the g-formula estimator with the IPW estimator of 
\cite{Barkley2020}
The same data generating process described above was repeated, 
with the exception of steps (iii) and (iv) where instead the treatment status for each individual was generated as a Bernoulli random variable with expectation 
$\pi_{ij} = \text{expit}(\rho_0+\rho_1L_{1i}+\rho_2L_{2i})$ 
where $\Brho=(\text{logit}(0.6), -0.01, -0.01)^\top$, 
and similarly, individual outcomes were generated as Bernoulli random variables with mean 
$\eta_{ij}=\text{expit}(\beta_0+\beta_1L_{1i}+\beta_2S_i+\beta_3L_{2i})$ 
where $\Bbeta=(\text{logit}(0.6), -0.01, -0.8, -0.01)^\top$. 
For each simulated dataset, the proposed g-formula estimator and the 
\cite{Barkley2020}
IPW estimator were computed. 
Results in the top panel of  Table \ref{sim table ipw} show both estimators are approximately unbiased, but the ESE of the IPW estimator is roughly twice as large. 
Because the IPW estimator is known to not perform well in the presence of large clusters (as in the bed net study), 
the simulation study was repeated, but in step (i), 
the number of individuals per cluster $N_i$ was simulated 
under two additional scenarios:
(i) $P(N_i = 20) =0.4, P(N_i = 50) = 0.35$, $P(N_i = 100) = 0.25$,
and
(ii) $P(N_i = 40) =0.4, P(N_i = 100) = 0.35$, $P(N_i = 200) = 0.25$. 
For these simulations, 
the bias of the IPW estimator increased as the cluster size increased,
and
the ESE was an order of magnitude larger than the g-formula estimator for most estimands; 
see the middle and bottom panels of  Table \ref{sim table ipw}.

\begin{center}
	\begin{table}[ht]
		\centering
		\caption{{Summary of simulation study results comparing g-formula and IPW as described in Section 3. Truth: true value of the estimand targeted by the estimator. Bias: average bias of the g-formula and IPW estimates over 1000 datasets. ESE: empirical standard error. Results are for all individuals (overall effect)}.}
		\label{sim table ipw}
		\begin{tabular}{cccccc}
            \hline
            Estimand & Truth & G-Formula Bias & G-Formula ESE & IPW Bias & IPW ESE \\ \hline
            \multicolumn{6}{c}{125 clusters of sizes 8, 16, 20} \\ \hline
            $\mu(\alpha=0.4)$ & 0.418 & -0.000 & 0.015 & -0.003 & 0.035 \\
            $\mu(\alpha=0.5)$ & 0.399 & 0.000 & 0.012 & -0.001 & 0.014 \\
            $\mu(\alpha=0.6)$ & 0.380 & 0.000 & 0.014 & -0.003 & 0.024 \\
            $\delta(\alpha=0.6, \alpha'=0.4)$ & -0.038 & 0.001 & 0.017 & 0.000 & 0.037 \\
            $\delta(\alpha=0.6, \alpha'=0.5)$ & -0.019 & 0.000 & 0.009 & -0.003 & 0.020 \\
            $\delta(\alpha=0.5, \alpha'=0.4)$ & -0.019 & 0.000 & 0.009 & 0.003 & 0.029 \\ \hline
            \multicolumn{6}{c}{125 clusters of sizes 20, 50, 100} \\ \hline
            $\mu(\alpha=0.4)$ & 0.418 & -0.000 & 0.011 & -0.006 & 0.247 \\
            $\mu(\alpha=0.5)$ & 0.399 & -0.000 & 0.007 & -0.002 & 0.016 \\
            $\mu(\alpha=0.6)$ & 0.380 & 0.000 & 0.011 & -0.008 & 0.266 \\
            $\delta(\alpha=0.6, \alpha'=0.4)$ & -0.038 & 0.000 & 0.018 & -0.002 & 0.362 \\
            $\delta(\alpha=0.6, \alpha'=0.5)$ & -0.019 & 0.000 & 0.009 & -0.006 & 0.262 \\
            $\delta(\alpha=0.5, \alpha'=0.4)$ & -0.019 & 0.000 & 0.009 & 0.004 & 0.240 \\ \hline
            \multicolumn{6}{c}{ 125 clusters of sizes 40, 100, 200} \\ \hline
             $\mu(\alpha=0.4)$ &  0.418 &  0.000&  0.010 &  -0.032 &  0.299 \\
             $\mu(\alpha=0.5)$ &  0.399 &  0.000&  0.005 &  -0.005 &  0.021 \\
             $\mu(\alpha=0.6)$ &  0.380 &  -0.000 &  0.010 &  -0.047 &  0.172 \\
             $\delta(\alpha=0.6, \alpha'=0.4)$ &  -0.038 &  -0.001 &  0.018 &  -0.015 &  0.339 \\
             $\delta(\alpha=0.6, \alpha'=0.5)$ &  -0.019 &  -0.000 &  0.009 &  -0.042 &  0.166 \\
             $\delta(\alpha=0.5, \alpha'=0.4)$ &  -0.019 &  -0.000 &  0.009 &   0.027 &  0.291 \\ \hline
        \end{tabular}
	\end{table}
\end{center}

\section{Analysis of Bed Net Use on Malaria in the Democratic Republic of the Congo} \label{Analysis}
\begin{figure}[ht]
	\centering
	\begin{subfigure}
		\centering
		\includegraphics[scale=0.48]{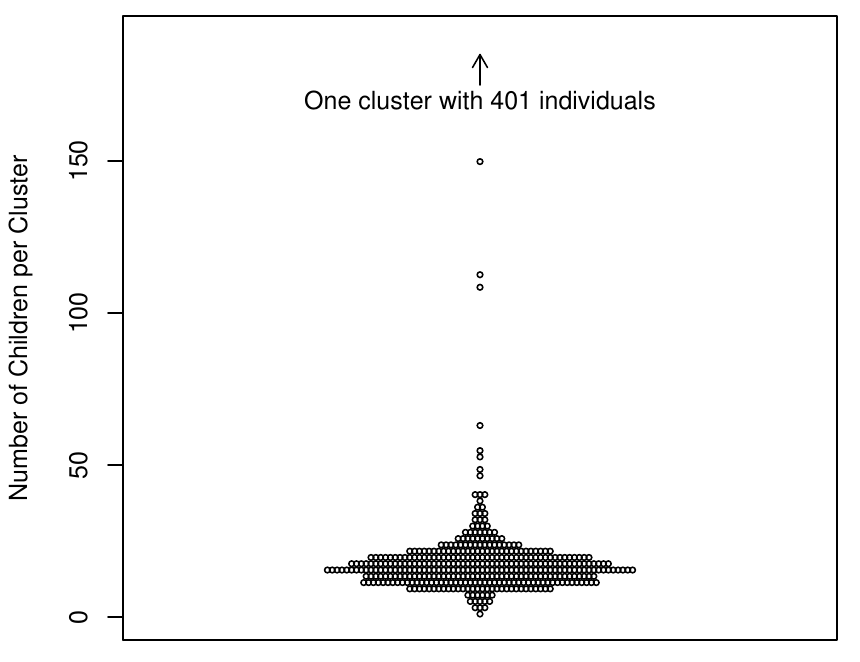}
	\end{subfigure}%
	\begin{subfigure}
		\centering
		\includegraphics[scale=0.483]{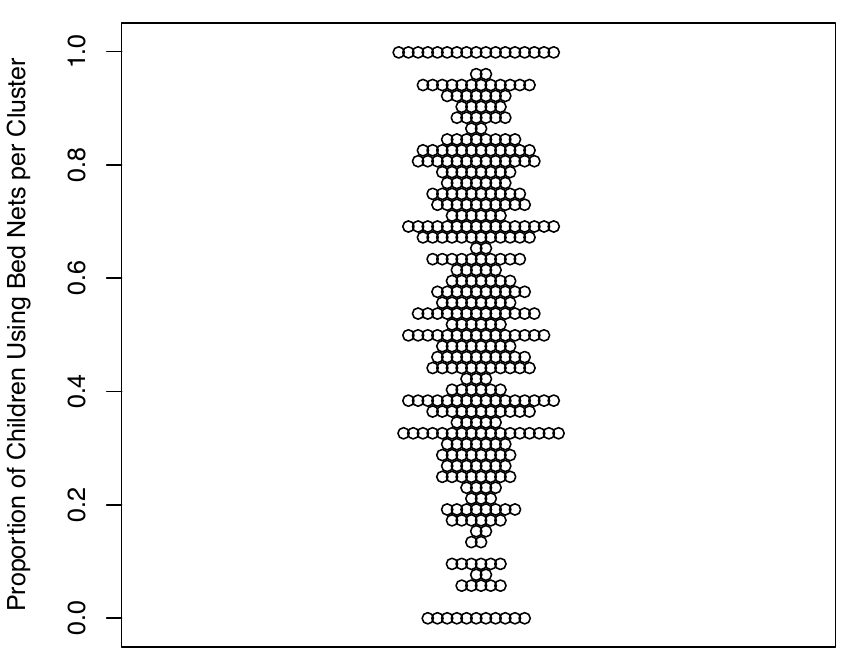}
	\end{subfigure}
	\caption{Malaria bed net study in the Democratic Republic of the Congo. Left panel: number of children with a measured malaria outcome per cluster. Right panel: proportion of children who used bed nets per cluster.}
	\label{fig:drc descriptives}
\end{figure}

The methods described above were applied to the DRC DHS survey to draw inference about the effects of bed nets on malaria in children when varying the proportion of children in this age range who use bed nets. As mentioned in Section \ref{Intro}, a single linkage agglomerative hierarchical cluster method \citep{Everitt2011} was used to group households of individuals into clusters. The maximum distance between any two households in the same cluster was constrained to not exceed 10 kilometers. This distance was selected based on the maximum flight distance of an \textit{Anopheles} mosquito \citep{Janko2018}. 
The GPS coordinates used in the clustering algorithm were randomly displaced from the actual location to prevent participant identification. Rural clusters were displaced up to 5 kilometers, while urban clusters were displaced up to 2 kilometers \citepalias{DHS}. 
Using this clustering algorithm, there were 395 clusters with at least one child that were not missing spatial information and other covariates. Figure \ref{fig:drc descriptives} displays the number of children per cluster, as well as the proportion of these children who used bed nets; on average, 55\% of children utilized bed nets. 

Because malaria was measured only in children, $Y$, $S$, and $N$ for each cluster were defined based only on children with a measured outcome.
Exchangeability was assumed conditional on the cluster-level proportion of women, 
as well as cluster-level averages of building materials (described below), urbanicity, altitude, age, temperature in the month of the survey, total precipitation in a 10-kilometer radius the month before the survey, and proportion of agricultural land cover within a 10-kilometer radius in 2013. 
The building material variable was defined similarly to \cite{Levitz2018}, where roof and wall materials were summed for each individual within a cluster. 
Natural materials were worth 0 points, rudimentary materials 1 point, and finished materials 2 points. 
Hence, for each individual, the building material variable was an integer between 0 and 4. 
The conditional exchangeability assumption supposes that, within strata of clusters defined by levels of these eight covariates, the proportion of children using bed nets is essentially randomized.

These covariates were selected based on prior research showing that each covariate is predictive of both the exposure and the outcome. 
Gender has been found to be associated with malaria; 
in particular, pregnant women may attract more malaria-carrying mosquitoes and have decreased malaria immunity 
\citep{Lindsay2000, Janko2018, Smith2021, Kuse2022, CDCmalaria}. 
Women, and in particular pregnant women, have been found to use bed nets more often \citep{ Olapeju2018, Kuse2022}. 
Age is associated with malaria prevalence in several studies, 
with children being particularly vulnerable due to low malaria immunity \citep{ Carneiro2010, Janko2018, Levitz2018, Smith2021, Kuse2022, CDCmalaria}. 
Children under five years old, along with older adults, tend to be more likely to use bed nets \citep{ Xu2014, Olapeju2018}. 
Housing quality and urbanicity are both associated with malaria prevalence, where traditional homes or sleeping outside are associated with increased malaria risk \citep{ Janko2018, Levitz2018, Smith2021}. 
Individuals sleeping in temporary housing such as tents or those in poor, rural areas
may not be able to afford bed nets or housing that prevents mosquito entry \citep{Smith2021, CDCmalaria},
while individuals in urban areas may be more inclined to use bed nets due to increased exposure to education and health systems \citep{Kuse2022}.
Altitude, temperature, precipitation, and agricultural land cover are all associated with malaria prevalence, 
with more favorable climates for malaria parasites and mosquito larvae leading to increased malaria risk \citep{ Janko2018, Levitz2018, Smith2021, CDCmalaria}. 
In areas where the perceived malaria risk is low due to unfavorable conditions for mosquitoes or malaria parasites, 
bed net use may be decreased \citep{ Xu2014}.
Additionally, in warmer temperatures, 
individuals may choose not to use bed nets if it’s too hot when using them or if they sleep outdoors \citep{ Xu2014, CDCmalaria}. 

In addition to conditional exchangeability, WSI was assumed conditional on the same set of covariates as the conditional exchangeability assumption. 
The WSI assumption supposes that within the strata of clusters having the same covariate values, 
the expected cluster-level malaria outcome depends only on the proportion of children using bed nets but not on which particular children use bed nets. 

\begin{figure}[ht]
	\centering
	\includegraphics[scale=0.8]{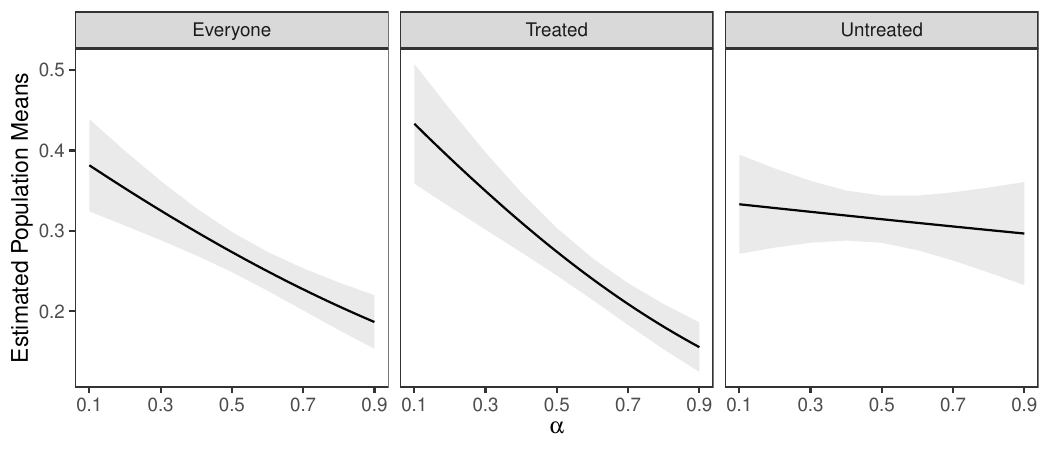}
	\caption{Estimates of the population mean estimands from the malaria bed net study. The proportion of treated children is denoted by policy $\alpha$. The shaded regions indicate point-wise 95\% confidence intervals.}
	\label{fig:popmeans}
\end{figure}

Figure \ref{fig:popmeans} displays g-formula estimates and 
corresponding point-wise 95\% confidence intervals 
of the population mean estimands over a range of policies $\alpha \in [0.1, 0.9]$ in all individuals, when treated, and when untreated. 
The left panel of Figure \ref{fig:popmeans} shows that the overall risk of malaria decreases as $\alpha$ increases, which is not surprising since bed nets are known to protect against malaria and bed net usage increases with $\alpha$. 
The middle panel of Figure \ref{fig:popmeans} demonstrates that the risk of malaria when treated also decreases as $\alpha$ increases. 
On the other hand, there appears to be little or no  effect when untreated (right panel of Figure \ref{fig:popmeans}).

\begin{figure}[ht]
	\centering
	\includegraphics[scale=0.8]{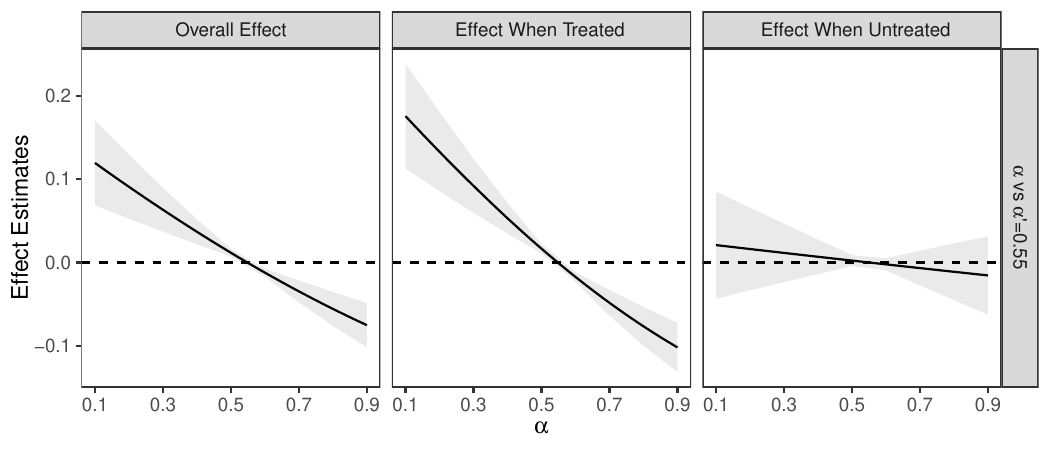}
	\caption{Estimated effects from the malaria bed net study. The proportion of treated children is denoted by policy $\alpha$. Effects contrast $\alpha$ with $\alpha '=0.55$, the current factual policy. The shaded regions indicate point-wise 95\% confidence intervals.}
	\label{fig:Chap 3 effects}
\end{figure}

Estimates and 
point-wise 95\% confidence intervals 
of the overall effects,  effects when treated, and  effects when untreated for different policies $\alpha$ compared to the current factual policy $\alpha'=0.55$ are displayed in Figure \ref{fig:Chap 3 effects}. These estimates approximate the expected change in the number of cases of malaria due to increasing or decreasing bed net use relative to current utilization. For example, $\hat{\delta}(\alpha=0.8, \alpha'=0.55)=-0.056$ ($95\%$ CI $-0.076, -0.035$) indicates that if 80\% of children in a cluster were to use bed nets, then we would expect 56 fewer cases of malaria per 1000 children on average. Similarly, for the  effect when treated, $\hat{\delta}(\alpha=0.8, \alpha'=0.55)=-0.077$ ($95\%$ CI $-0.10, -0.054$), indicating we would expect 77 fewer cases of malaria per 1000 treated children on average if 80\% of children in a cluster were to use bed nets. On the other hand, the  effect when untreated for $\alpha=0.8$ compared to $\alpha'=0.55$ is $-0.011$ ($95\%$ CI $-0.045, 0.023$), suggesting no or modest benefit of increasing bed net use to non-users. 

For the sake of comparison, the \cite{Barkley2020}
IPW estimator was also applied to the DRC DHS data to estimate the bed net effects. However, the mixed effects model used to estimate the group propensity scores did not converge, hence it was not possible to compute the IPW estimates. Given that the DRC data includes several large clusters, it is not surprising issues were encountered when attempting to compute the IPW estimator. 
See \cite{Saul2017} for further discussion related to computational issues of partial interference IPW estimators. A possible workaround would be to exclude the large clusters \citep{Chakladar2022}, but this would inefficiently discard data and limit the generalizability of the results. 

\begin{figure}[ht]
	\centering
	\includegraphics[scale=0.8]{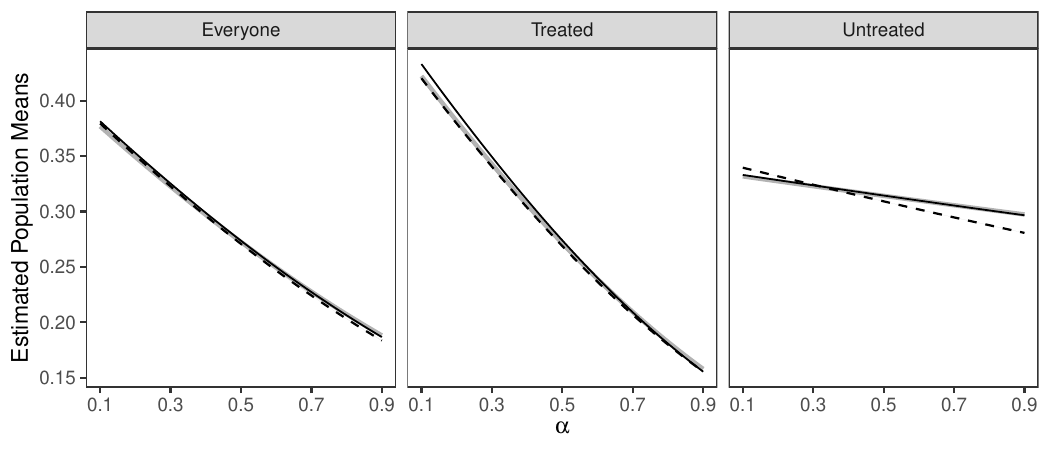}
	\caption{Estimates of the population mean estimands from the malaria bed net study. The proportion of treated children is denoted by policy $\alpha$. Solid black lines represent 10 km, solid gray lines represent 5 km, and dashed lines represent 2.5 km clusters.}
	\label{fig:sense}
\end{figure}

The results above are based on the clustering of households  such that the maximum distance between any two households in the same cluster was 10 km. Sensitivity analyses were performed, where clusters were instead defined based on maximum distances of 5 km and 2.5 km. There were 415 clusters in the 5 km analysis and 449 clusters in the 2.5 km analysis that were not missing spatial information and had at least one child. Population mean estimates were very similar between the 2.5 km, 5 km, and 10 km analyses; see  Figure \ref{fig:sense}.

The g-formula approach relies on the correct specification of the outcome model,
and thus,
assessing the robustness of results to model specification is essential when using this method in practice.
Various sensitivity analyses of the DRC data did not result in meaningful changes in the conclusions. For example, the results in Supplementary Figures \ref{fig: sensitivity drop individual covariates1} and \ref{fig: sensitivity drop individual covariates2} show the population mean estimates were similar to the original estimates when each of the eight covariates was individually dropped from the outcome model. 
The sensitivity of the results to the specified outcome regression link function was also assessed. 
Results in Supplementary Figure \ref{fig:sensitivity link and interaction} show that when the probit link function was used instead of the logit link function for treatment and outcome models,
the estimates were almost identical to the original scenario. 
Finally,
an interaction term between $S$ and the average total precipitation in a 10 km radius at the month before the survey was included in the outcome model,
and the estimates did not appreciably change relative to the original estimates from the model without the interaction term (Supplementary Figure \ref{fig:sensitivity link and interaction}).

\begin{figure}[h]
	\centering
	\includegraphics[scale=0.8]{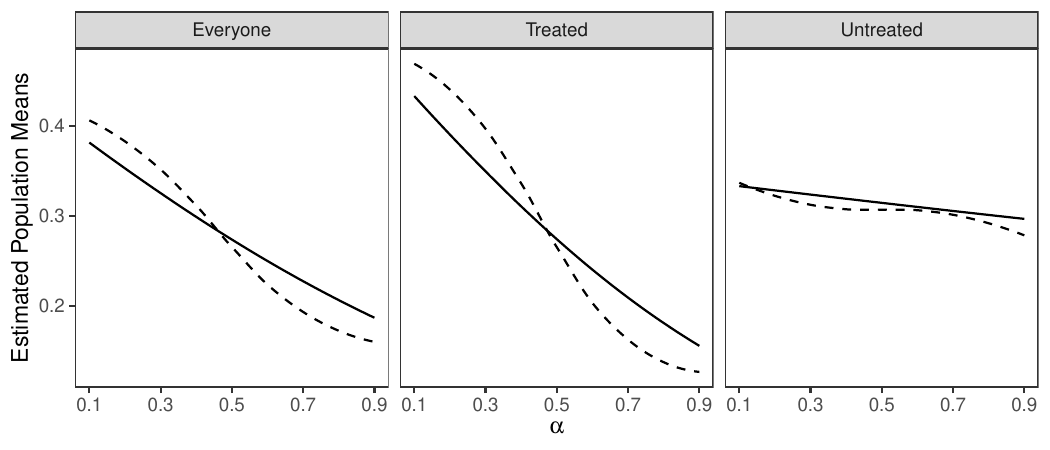}
	\caption{Estimates of the population mean estimands from the malaria bed net study for the children-only policy (solid lines) and general population policy (dashed lines).}
	\label{fig:subgroups}
\end{figure}

To investigate the effect of changing the proportion of the entire population who use bed nets, the 10-kilometer clusters were also analyzed using the methods from Section \ref{subgroups} with separate parametric models fit for $S_1$ given $\mathbf{L}, S_2$ and for $S_2$ given $\mathbf{L}.$ The estimated population means for the general population policy compared to the children-only policy are shown in  Figure \ref{fig:subgroups}. Changes in the general population policy are associated with greater changes in the mean outcome in all individuals and when treated compared to the children-only policy. However, the largest difference in estimated population means between the general population policy and the children-only policy is only 0.05. For the  effect when untreated, the estimates are approximately the same for both the children-only and general population policies.  

\section{Discussion} \label{Discussion}
In the presence of partial interference, the proposed g-formula estimator is an alternative to existing IPW estimators, 
such as those proposed in \cite{Tchetgen2012}. 
The g-formula estimator can accommodate large clusters, unlike IPW estimators \citep{Liu2019,Chakladar2022}, 
and does not suffer from the g-null paradox that may occur in the absence of interference. 
Like the IPW estimators of 
\cite{Papadogeorgou2019} and \cite{Barkley2020},
the proposed methods target counterfactual estimands, which allow for within-cluster dependence of treatment selection and thus may be more relevant to policymakers. 
While motivated by infectious disease prevention studies, the g-formula methods developed in this paper are applicable in other settings where partial interference may be present.

The approach used in this paper requires only cluster-level covariates, exposures, and outcomes, and thus may be of particular utility in settings where individual-level data is not available and only cluster-level variables are observed. 
As noted by one reviewer, the approach in this paper accommodates partial interference by essentially reframing as a `classic causal inference' problem where the clusters as units and i.i.d. unit-level data are observed. 
Moreover, the g-formula estimator could easily be adapted to the no interference setting where the exposure is a fraction between 0 and 1.

The analysis of the DRC Demographic and Health Survey data in Section \ref{Analysis} provides evidence that increasing the proportion of children who use bed nets reduces the proportion of children with malaria. The protective effect of bed nets was found to be more pronounced in children who use bed nets compared to children who do not. These results complement existing research on the population-level effects of bed nets. For example, community-level effects of bed net use have previously been found to affect the overall risk of malaria among individuals within the community \citep{Binka1996, Hii2001}. Additionally, previous studies have shown that community-level bed net coverage can have indirect (i.e., spillover) protective effects \citep{Binka1998, Howard2000, Maxwell2002, Hawley2003}. Similar to the results in this paper, \cite{Escamilla2017} found that the protective association of community-level bed net coverage with malaria morbidity in Malawi was greater among children who used bed nets compared to children who did not. Likewise, using the same DRC data as in this paper, \cite{Levitz2018} found community-level bed net use was associated with a lower prevalence of malaria, and in individuals who used bed nets, the odds of malaria were further diminished. The DRC data analysis presented in Section \ref{Analysis} extends these previous association-type analyses of population-level bed net usage to inference about the different causal effects of bed nets. Such inference quantifies the expected number of malaria cases prevented for different levels of bed net coverage when individuals use bed nets as well as when individuals do not, thus providing a clearer interpretation of and additional insight into the population-level impact of bed nets for investigators and policymakers.

There are several possible areas of further methodological research related to this paper. 
For example, the consistency of the proposed g-formula estimator requires that the parametric models be correctly specified.
These models make explicit assumptions on the cluster-level data generating process (DGP) without explicitly imposing any assumptions on the individual-level DGP.  
That said, individual-level DGPs that do not marginalize to the assumed cluster-level DGP are implicitly assumed to not hold. 
Future research could explore relaxing these parametric assumptions, 
perhaps by using semiparametric or nonparametric (i.e., machine learning) methods. 
In no interference setting, \cite{kennedy19} considers estimands similar to the estimands in this paper corresponding to average potential outcomes when the proportion treated is increased.
Kennedy proposes nonparametric doubly robust estimators, which tend to have smaller bias 
but larger variance than the parametric approach considered in this paper. 
Extensions of the proposed method which relax the partial interference assumption and 
thereby allowing for more general (e.g., spatial) interference \citep{wang2020design, leung2022rate} could be considered. 
The method developed here considers only a univariate exposure, such as bed net usage in the malaria study, and makes no assumptions about other possible exposures. Future research could extend the methods in this paper to draw inference about the joint effects of multivariate exposures in the presence of partial interference. In the context of malaria, such methods could be used to assess the effects of policies on bed net usage and other risk factors for malaria, such as spraying of insecticides or other approaches to mosquito control. The RTS,S/AS01 malaria vaccine was recently recommended for widespread use by the World Health Organization \citep{Maxmen2021}, and it would also be of interest to assess the joint effect of bed nets and vaccination. 
As the total number of bed nets may be limited in practice, 
it may be helpful to extend the proposed methods to estimate
an optimal bed net allocation strategy based on cluster characteristics, 
subject to some constraints on the total number of bed nets available \citep{ananth2020optimal}.
Finally, the approach in this paper defines the unit as the cluster, which has advantages but does not easily permit inference about the direct effect. G-formula based estimates of direct effects could be developed, presumably based on individual-level potential outcomes.

\section*{Acknowledgements}

The authors thank Shaina Alexandria, Bryan Blette, M.\ Elizabeth Halloran, Sam Rosin, Bonnie Shook-Sa, and Jaffer Zaidi for providing comments on the manuscript. The authors also thank Mark Janko for providing temperature, precipitation, and agricultural density data. This work was partially supported by NIH grants R01 AI085073 and T32 ES007018.\vspace*{-8pt}

\section*{Data Availability Statement}
The DRC survey data is available upon request at \url{http://www.dhsprogram.com}, and the corresponding spatial data is available at \url{http://spatialdata.dhsprogram.com}.

\section*{Supporting Information}
R code to replicate the simulation study is available at \url{https://github.com/KilpatrickKW}. 

\bibliographystyle{plainnat}
\bibliography{GFormulaRef}       % Bibliography file (usually '*.bib')

\begin{thebibliography}{45}
\providecommand{\natexlab}[1]{#1}
\providecommand{\url}[1]{\texttt{#1}}
\expandafter\ifx\csname urlstyle\endcsname\relax
  \providecommand{\doi}[1]{doi: #1}\else
  \providecommand{\doi}{doi: \begingroup \urlstyle{rm}\Url}\fi

\bibitem[Ananth(2020)]{ananth2020optimal}
Abhishek Ananth.
\newblock Optimal treatment assignment rules on networked populations.
\newblock Technical report, Cornell University, 2020.

\bibitem[Barkley et~al.(2020)Barkley, Hudgens, Clemens, Ali, and Emch]{Barkley2020}
B.~G. Barkley, M.~G. Hudgens, J.~D. Clemens, M.~Ali, and M.~E. Emch.
\newblock Causal inference from observational studies with clustered interference, with application to a cholera vaccine study.
\newblock \emph{Annals of Applied Statistics}, 14\penalty0 (3):\penalty0 1432--1448, 2020.

\bibitem[Binka et~al.(1996)Binka, Kubaje, Adjuik, Williams, Lengeler, Maude, Armah, Kajihara, Adiamah, and Smith]{Binka1996}
F.~N. Binka, A.~Kubaje, M.~Adjuik, L.~A. Williams, C.~Lengeler, G.~H. Maude, G.~E. Armah, B.~Kajihara, J.~H. Adiamah, and P.~G. Smith.
\newblock Impact of permethrin impregnated bednets on child mortality in kassena-nankana district, ghana: a randomized controlled trial.
\newblock \emph{Tropical Medicine \& International Health}, 1\penalty0 (2):\penalty0 147--154, 1996.

\bibitem[Binka et~al.(1998)Binka, Indome, and Smith]{Binka1998}
F.~N. Binka, F.~Indome, and T.~Smith.
\newblock Impact of spatial distribution of permethrin-impregnated bed nets on child mortality in rural northern ghana.
\newblock \emph{The American Journal of Tropical Medicine and Hygiene}, 59\penalty0 (1):\penalty0 80--85, 1998.

\bibitem[Carneiro et~al.(2010)Carneiro, Roca-Feltrer, Griffin, Smith, Tanner, Schellenberg, Greenwood, and Schellenberg]{Carneiro2010}
I.~Carneiro, A.~Roca-Feltrer, J.T. Griffin, L.~Smith, M.~Tanner, J.A. Schellenberg, B.~Greenwood, and D.~Schellenberg.
\newblock Age-patterns of malaria vary with severity, transmission intensity and seasonality in sub-{S}aharan {A}frica: a systematic review and pooled analysis.
\newblock \emph{PLOS ONE}, 5\penalty0 (2):\penalty0 1--10, 02 2010.

\bibitem[{CDC: Center for Disease Control and Prevention}(2023)]{CDCmalaria}
{CDC: Center for Disease Control and Prevention}.
\newblock Parasites-malaria.
\newblock \url{https://www.cdc.gov/parasites/malaria/index.html}, 2023.

\bibitem[Chakladar et~al.(2022)Chakladar, Hudgens, Halloran, Clemens, Ali, and Emch]{Chakladar2022}
S.~Chakladar, M.~G. Hudgens, M.~E. Halloran, J.~D. Clemens, M.~Ali, and M.~E. Emch.
\newblock Inverse probability weighted estimators of vaccine effects accommodating partial interference and censoring.
\newblock \emph{Biometrics}, 78:\penalty0 777--788, 2022.

\bibitem[Cox(1958)]{Cox1958}
D.~R. Cox.
\newblock \emph{Planning of Experiments}.
\newblock New York: Wiley, 1958.

\bibitem[Crawford et~al.(2019)Crawford, Morozova, Buchanan, and Spiegelman]{Crawford2019}
F.~W. Crawford, O.~Morozova, A.~L. Buchanan, and D.~Spiegelman.
\newblock Interpretation of the individual effect under treatment spillover.
\newblock \emph{American Journal of Epidemiology}, 188\penalty0 (8):\penalty0 1407--1409, 2019.

\bibitem[Escamilla et~al.(2017)Escamilla, Alker, Dandalo, Juliano, Miller, Kamthuza, Tembo, Tegha, Martinson, Emch, and Hoffman]{Escamilla2017}
V.~Escamilla, A.~Alker, L.~Dandalo, J.~J. Juliano, W.~C. Miller, P.~Kamthuza, T.~Tembo, G.~Tegha, F.~Martinson, M.~Emch, and I.~F. Hoffman.
\newblock Effects of community-level bed net coverage on malaria morbidity in lilongwe, malawi.
\newblock \emph{Malaria Journal}, 16\penalty0 (142):\penalty0 1--9, 2017.

\bibitem[Everitt et~al.(2011)Everitt, Landau, Leese, and Stahl]{Everitt2011}
B.~S. Everitt, S.~Landau, M.~Leese, and D.~Stahl.
\newblock \emph{Cluster Analysis}.
\newblock John Wiley, {F}ifth edition, 2011.

\bibitem[Hawley et~al.(2003)Hawley, Phillips-Howard, Ter~Kuile, Terlouw, Vulule, Ombok, Nahlen, Gimnig, Kariuki, Kolczak, and Hightower]{Hawley2003}
W.~A. Hawley, P.~A. Phillips-Howard, F.~O. Ter~Kuile, D.~J. Terlouw, J.~M. Vulule, M.~Ombok, B.~L. Nahlen, J.~E. Gimnig, S.~K. Kariuki, M.~S. Kolczak, and A.~W. Hightower.
\newblock Community-wide effects of permethrin-treated bed nets on child mortality and malaria morbidity in western kenya.
\newblock \emph{The American Journal of Tropical Medicine and Hygiene}, 68\penalty0 (4\_suppl):\penalty0 121--127, 2003.

\bibitem[Hern{\'{a}}n and Robins(2006)]{Hernan2006}
M.~A. Hern{\'{a}}n and J.~M. Robins.
\newblock {Estimating causal effects from epidemiological data}.
\newblock \emph{Journal of Epidemiology {\&} Community Health}, 60\penalty0 (7):\penalty0 578--586, 2006.

\bibitem[Hii et~al.(2001)Hii, Smith, Vounatsou, Alexander, Mai, Ibam, and Alpers]{Hii2001}
J.~L.~K. Hii, T.~Smith, P.~Vounatsou, N.~Alexander, A.~Mai, E.~Ibam, and M.~P. Alpers.
\newblock Area effects of bednet use in a malaria-endemic area in papua new guinea.
\newblock \emph{Transactions of the Royal Society of Tropical Medicine and Hygiene}, 95\penalty0 (1):\penalty0 7--13, 2001.

\bibitem[Howard et~al.(2000)Howard, Omumbo, Nevill, Some, Donnelly, and Snow]{Howard2000}
S.~C. Howard, J.~Omumbo, C.~Nevill, E.~S. Some, C.~A. Donnelly, and R.~W. Snow.
\newblock Evidence for a mass community effect of insecticide-treated bednets on the incidence of malaria on the kenyan coast.
\newblock \emph{Transactions of the Royal Society of Tropical Medicine and Hygiene}, 94\penalty0 (4):\penalty0 357--360, 2000.

\bibitem[Hudgens and Halloran(2008)]{Hudgens2008}
M.~G. Hudgens and M.~E. Halloran.
\newblock Toward causal inference with interference.
\newblock \emph{Journal of the American Statistical Association}, 103\penalty0 (482):\penalty0 832--842, 2008.

\bibitem[Janko et~al.(2018)Janko, Irish, Reich, Peterson, Doctor, Mwandagalirwa, Likwela, Tshefu, Meshnick, and Emch]{Janko2018}
M.~M. Janko, S.~R. Irish, B.~J. Reich, M.~Peterson, S.~M. Doctor, M.~K. Mwandagalirwa, J.~L. Likwela, A.~K. Tshefu, S.~R. Meshnick, and M.~E. Emch.
\newblock The links between agriculture, \emph{{A}nopheles} mosquitoes, and malaria risk in children younger than 5 years in the {D}emocratic {R}epublic of the {C}ongo: a population-based, cross-sectional, spatial study.
\newblock \emph{The Lancet Planetary Health}, 2\penalty0 (2):\penalty0 e74--e82, 2018.

\bibitem[Kennedy(2019)]{kennedy19}
Edward~H. Kennedy.
\newblock Nonparametric causal effects based on incremental propensity score interventions.
\newblock \emph{Journal of the American Statistical Association}, 114\penalty0 (526):\penalty0 645--656, 2019.

\bibitem[Kuse et~al.(2022)Kuse, Chikako, Bacha, Hagan~Jr, Seidu, and Ahinkorah]{Kuse2022}
K.A. Kuse, T.U. Chikako, R.H. Bacha, J.E. Hagan~Jr, A.-A. Seidu, and B.O. Ahinkorah.
\newblock Multilevel modelling of individual, community and regional level factors associated with insecticide-treated net usage among pregnant women in {E}thiopia.
\newblock \emph{Healthcare}, 10\penalty0 (8), 2022.

\bibitem[Lee et~al.(2023)Lee, Zeng, and Hudgens]{lee2023efficient}
Chanhwa Lee, Donglin Zeng, and Michael~G Hudgens.
\newblock Efficient nonparametric estimation of stochastic policy effects with clustered interference.
\newblock \emph{arXiv:2212.10959v2}, 2023.

\bibitem[Leung(2022)]{leung2022rate}
Michael~P Leung.
\newblock Rate-optimal cluster-randomized designs for spatial interference.
\newblock \emph{The Annals of Statistics}, 50\penalty0 (5):\penalty0 3064--3087, 2022.

\bibitem[Levitz et~al.(2018)Levitz, Janko, Mwandagalirwa, Thwai, Likwela, Tshefu, Emch, and Meshnick]{Levitz2018}
L.~Levitz, M.~Janko, K.~Mwandagalirwa, K.~L. Thwai, J.~L. Likwela, A.~K. Tshefu, M.~Emch, and S.~R. Meshnick.
\newblock Effect of individual and community-level bed net usage on malaria prevalence among under-fives in the {D}emocratic {R}epublic of {C}ongo.
\newblock \emph{Malaria Journal}, 17\penalty0 (1):\penalty0 39, 2018.

\bibitem[Lindsay et~al.(2000)Lindsay, Ansell, Selman, Cox, Hamilton, and Walraven]{Lindsay2000}
S.~Lindsay, J.~Ansell, C.~Selman, V.~Cox, K.~Hamilton, and G.~Walraven.
\newblock Effect of pregnancy on exposure to malaria mosquitoes.
\newblock \emph{The Lancet}, 355\penalty0 (9219):\penalty0 1972, 2000.

\bibitem[Liu et~al.(2019)Liu, Hudgens, Saul, Clemens, Ali, and Emch]{Liu2019}
L.~Liu, M.~G. Hudgens, B.~Saul, J.~D. Clemens, M.~Ali, and M.~E. Emch.
\newblock Doubly robust estimation in observational studies with partial interference.
\newblock \emph{Stat}, 8\penalty0 (1):\penalty0 e214, 2019.

\bibitem[Maxmen(2021)]{Maxmen2021}
A.~Maxmen.
\newblock Scientists hail historic malaria vaccine approval-but point to challenges ahead.
\newblock \emph{Nature}, 2021.
\newblock DOI: 10.1038/d41586-021-02755-5.

\bibitem[Maxwell et~al.(2002)Maxwell, Msuya, Sudi, Njunwa, Carneiro, and Curtis]{Maxwell2002}
C.~A. Maxwell, E.~Msuya, M.~Sudi, K.~J. Njunwa, I.~A. Carneiro, and C.~F. Curtis.
\newblock Effect of community-wide use of insecticide-treated nets for 3--4 years on malarial morbidity in tanzania.
\newblock \emph{Tropical Medicine and International Health}, 7\penalty0 (12):\penalty0 1003--1008, 2002.

\bibitem[{Min}(2014)]{DHS}
\emph{R\'{e}publique {D}\'{e}mocratique du {C}ongo {E}nqu\^{e}te {D}\'{e}mographique et de {S}ant\'{e} ({EDS-RDC}) 2013-2014 [Dataset]. CDPR61SD, CDGE61FL.}
\newblock {Minist\'{e}re du Plan et Suivi de la Mise en oeuvre de la R\'{e}volution de la Modernit\'{e} (MPSMRM), Minist\'{e}re de la Sant\'{e} Publique (MSP), and ICF International}, Rockville, Maryland, USA, 2014.
\newblock Rockville, Maryland, USA: MPSMRM, MSP and ICF International [Producers], ICF [Distributer].

\bibitem[Mu{\~n}oz and Van Der~Laan(2012)]{munoz12}
Iv{\'a}n~D{\'\i}az Mu{\~n}oz and Mark Van Der~Laan.
\newblock Population intervention causal effects based on stochastic interventions.
\newblock \emph{Biometrics}, 68\penalty0 (2):\penalty0 541--549, 2012.

\bibitem[Olapeju et~al.(2018)Olapeju, Choiriyyah, Lynch, Acosta, Blaufuss, Filemyr, Harig, Monroe, Selby, Kilian, and Koenker]{Olapeju2018}
B.~Olapeju, I.~Choiriyyah, M.~Lynch, A.~Acosta, S.~Blaufuss, E.~Filemyr, H.~Harig, A.~Monroe, R.A. Selby, A.~Kilian, and H.~Koenker.
\newblock Age and gender trends in insecticide-treated net use in sub-{S}aharan {A}frica: a multi-country analysis.
\newblock \emph{Malaria Journal}, 17\penalty0 (423):\penalty0 1--12, 2018.

\bibitem[Papadogeorgou et~al.(2019)Papadogeorgou, Mealli, and Zigler]{Papadogeorgou2019}
G.~Papadogeorgou, F.~Mealli, and C.~M. Zigler.
\newblock Causal inference with interfering units for cluster and population level treatment allocation programs.
\newblock \emph{Biometrics}, 75\penalty0 (3):\penalty0 778--787, 2019.

\bibitem[Park and Kang(2022)]{park2022efficient}
Chan Park and Hyunseung Kang.
\newblock Efficient semiparametric estimation of network treatment effects under partial interference.
\newblock \emph{Biometrika}, 109\penalty0 (4):\penalty0 1015--1031, 2022.

\bibitem[Richardson and Robins(2013)]{Richardson2013}
T.~S. Richardson and J.~M. Robins.
\newblock Single world intervention graphs (swigs): A unification of the counterfactual and graphical approaches to causality.
\newblock \emph{Center for the Statistics and the Social Sciences, University of Washington Series. Working Paper}, 128\penalty0 (30):\penalty0 2013, 2013.

\bibitem[Robins(1986)]{Robins1986}
J.~Robins.
\newblock {A new approach to causal inference in mortality studies with sustained exposure period — application to control of the healthy worker survivor effect}.
\newblock \emph{Mathematical Modelling}, 7:\penalty0 1393--1512, 1986.
\newblock ISSN 0270-0255.

\bibitem[Robins and Wasserman(1997)]{Robins1997proceedings}
J.~M. Robins and L.~Wasserman.
\newblock Estimation of effects of sequential treatments by reparameterizing directed acyclic graphs.
\newblock \emph{Proceedings of the Thirteenth Conference on Uncertainty in Artificial Intelligence}, 1997.

\bibitem[Rubin(2005)]{Rubin2005}
D.~B. Rubin.
\newblock Causal inference using potential outcomes: Design, modeling, decisions.
\newblock \emph{Journal of the American Statistical Association}, 100\penalty0 (469):\penalty0 322--331, 2005.

\bibitem[Saul and Hudgens(2017)]{Saul2017}
B.~C. Saul and M.~G. Hudgens.
\newblock {A recipe for inferference: Start with causal inference. Add interference. Mix well with R}.
\newblock \emph{Journal of Statistical Software}, 82\penalty0 (2), 2017.

\bibitem[Smith et~al.(2021)Smith, Mumbengegwi, Haindongo, Cueto, Roberts, Gosling, Uusiku, Kleinschmidt, Bennett, and Sturrock]{Smith2021}
J.L. Smith, D.~Mumbengegwi, E.~Haindongo, C.~Cueto, K.W. Roberts, R.~Gosling, P.~Uusiku, I.~Kleinschmidt, A.~Bennett, and H.J. Sturrock.
\newblock Malaria risk factors in northern {N}amibia: the importance of occupation, age and mobility in characterizing high-risk populations.
\newblock \emph{PLOS ONE}, 16\penalty0 (6):\penalty0 1--23, 06 2021.

\bibitem[Sobel(2006)]{Sobel2006}
M.~E. Sobel.
\newblock What do randomized studies of housing mobility demonstrate? {C}ausal inference in the face of interference.
\newblock \emph{Journal of the American Statistical Association}, 101\penalty0 (476):\penalty0 1398--1407, 2006.

\bibitem[Stefanski and Boos(2002)]{Stefanski2002}
L.~A. Stefanski and D.~D. Boos.
\newblock The calculus of {M}-estimation.
\newblock \emph{The American Statistician}, 56\penalty0 (1):\penalty0 29--38, 2002.

\bibitem[Taubman et~al.(2009)Taubman, Robins, Mittleman, and Hern{\'a}n]{Taubman2009}
S.~L. Taubman, J.~M. Robins, M.~A. Mittleman, and M.~A. Hern{\'a}n.
\newblock Intervening on risk factors for coronary heart disease: an application of the parametric g-formula.
\newblock \emph{International Journal of Epidemiology}, 38\penalty0 (6):\penalty0 1599--1611, 2009.

\bibitem[Tchetgen~Tchetgen and VanderWeele(2012)]{Tchetgen2012}
E.~J. Tchetgen~Tchetgen and T.~J. VanderWeele.
\newblock On causal inference in the presence of interference.
\newblock \emph{Statistical Methods in Medical Research}, 21\penalty0 (1):\penalty0 55--75, 2012.

\bibitem[VanderWeele and Tchetgen~Tchetgen(2011)]{VanderWeele2011}
T.~J. VanderWeele and E.~J. Tchetgen~Tchetgen.
\newblock Effect partitioning under interference in two-stage randomized vaccine trials.
\newblock \emph{Statistics \& Probability Letters}, 81\penalty0 (7):\penalty0 861--869, 2011.

\bibitem[Wang et~al.(2020)Wang, Samii, Chang, and Aronow]{wang2020design}
Ye~Wang, Cyrus Samii, Haoge Chang, and PM~Aronow.
\newblock Design-based inference for spatial experiments with interference.
\newblock \emph{arXiv preprint arXiv:2010.13599}, 2020.

\bibitem[Wen et~al.(2023)Wen, Marcus, and Young]{wen23}
Lan Wen, Julia~L Marcus, and Jessica~G Young.
\newblock Intervention treatment distributions that depend on the observed treatment process and model double robustness in causal survival analysis.
\newblock \emph{Statistical Methods in Medical Research}, 32\penalty0 (3):\penalty0 509--523, 2023.

\bibitem[Xu et~al.(2014)Xu, Liao, Liu, Nie, and Havumaki]{Xu2014}
J.-W. Xu, Y.-M. Liao, H.~Liu, R.-H. Nie, and J.~Havumaki.
\newblock Use of bed nets and factors that influence bed net use among {J}inuo ethnic minority in southern {C}hina.
\newblock \emph{PLOS ONE}, 9\penalty0 (7):\penalty0 e103780, 07 2014.

\end{thebibliography}

\newpage

\section*{Appendix}

\subsection*{A.1 Large sample properties of the proposed estimators}

The g-formula estimators in Section \ref{estimators} of the main paper can be shown to be consistent and asymptotically Normal using standard large-sample estimating equation theory. 
Let 
$\Btheta
=
(\Brho, \gamma_{0\alpha}, \gamma_{0\alpha'}, \Bbeta, 
\mu(\alpha), \mu(\alpha'), \delta(\alpha, \alpha'))^\top$. 
Estimating functions for $\hat{\Brho}$ and $\hat{\Bbeta}$ are given by score equations corresponding to the binomial models $P(S|\mathbf{L};\Brho)$ and $P(Y|S, \mathbf{L};\Bbeta)$. 
Denote these score equations by 
$\psi_{\Brho}(O;\Btheta)$ and $\psi_{\Bbeta}(O;\Btheta)$. 
For policy $\alpha$, let 
$\psi_{\gamma_{0\alpha}}(O;\Btheta)
=
E_\alpha(S|\mathbf{L};\gamma_{0\alpha},\Brho_1)
-
\alpha$ 
where 
$E_\alpha(S|\mathbf{L};\gamma_{0\alpha},\Brho_1)
=
\text{expit}(\gamma_{0\alpha}+\Brho_1^\top\mathbf{L})$, 
and let
\begin{equation*}
    \psi_{\mu(\alpha)}(O;\Btheta)
    =
    \sum_{s \in \mathcal{S}} 
        E(Y|S=s,\mathbf{L};\Bbeta)
        P_\alpha(S=s|\mathbf{L};\gamma_{0\alpha},\Brho_1)
    -
    \mu(\alpha).
\end{equation*}
Define 
$\psi_{\delta(\alpha, \alpha')}(O;\Btheta)
=
\psi_{\mu(\alpha)}(O;\Btheta)
-
\psi_{\mu(\alpha')}(O;\Btheta)$, 
and let 
$\psi_{\Btheta}
=
(\psi_{\Brho}, \psi_{\gamma_{0\alpha}}, \psi_{\gamma_{0\alpha'}},
\psi_{\Bbeta}, \psi_{\mu(\alpha)}, \psi_{\mu(\alpha')}$,
$\psi_{\delta(\alpha, \alpha')})^\top$. 
Then the estimator 
$\hat{\Btheta}
=
(\hat{\Brho}, \hat{\gamma}_{0\alpha}, \hat{\gamma}_{0\alpha'}, 
\hat{\Bbeta}, \hat{\mu}(\alpha), \hat{\mu}(\alpha'), 
\hat{\delta}(\alpha, \alpha'))^\top$ 
is the solution to the vector estimating equation 
$\sum_{i=1}^m \psi_{\theta}(O;\Btheta)=\mathbf{0}$.

It is straightforward to show these estimating equations are unbiased. 
Because $\psi_{\Brho}(O;\Btheta)$ and $\psi_{\Bbeta}(O;\Btheta)$ are score equations, 
$\int \psi_{\Brho}(O;\Btheta)dF_O(O)=0$ 
and $\int \psi_{\Bbeta}(O;\Btheta)dF_O(O)=0$ 
where $F_O(O)$ denotes the distribution of the observed variables $O$. 
For policy $\alpha$, $\gamma_{0\alpha}$ is the solution to (5), implying $E\{\psi_{\gamma_{0\alpha}}(O;\Btheta)\} = 0$. 
Next note
\begin{align*}
    E\{\psi_{\mu(\alpha)}(O;\Btheta)\}
    &=
    E\left\{
        \sum_{s \in \cSN} 
            E(Y|S=s,\mathbf{L})
            P_\alpha(S=s|\mathbf{L})
    \right\}
    -
    \mu(\alpha)
    \\
    &=
    \int_{\Bl}
        \sum_{s \in \cSn}
            E \big( Y | S = s, \BL = \Bl \big) 
            P_{\alpha} \big( S = s | \BL = \Bl \big)
    d F_{\BL}(\Bl)
    -
    \mu(\alpha)
    \\
	&=
    0
\end{align*}
assuming $Y|S,\mathbf{L}$ and $S|\mathbf{L}$ models are correctly specified.

From standard large-sample estimating equation theory, 
it follows that under suitable regularity conditions, 
$\hat{\Btheta} \rightarrow_p \Btheta$ 
and $\sqrt{m}(\hat{\Btheta} - \Btheta) \rightarrow_d N(0,\Sigma)$ 
where $\Sigma=U^{-1}W(U^{-1})^\top$ for 
$U=E\{-\dot{\psi}_{\Btheta}(O;\Btheta)\}$,
where 
$\dot{\psi}_{\Btheta}(O;\Btheta)
=
\partial \psi_{\Btheta}(O;\Btheta)
/
\partial \Btheta^\top$,  
and 
$W
=
E\{\psi_{\Btheta}(O;\Btheta)^{\otimes 2}\}$. 
The asymptotic variance $\Sigma$ can be consistently estimated by the empirical sandwich variance estimator
$\widehat{\Sigma}
=
\widehat{U}^{-1} \widehat{W}(\widehat{U}^{-1})^\top$ 
where 
$\widehat{U}=m^{-1}\sum_{i=1}^m -\dot{\psi}_{\Btheta}(O_i;\hat{\Btheta})$ and 
$\widehat{W}= m^{-1}\sum_{i=1}^m \psi_{\Btheta}(O_i;\hat{\Btheta})^{\otimes 2}$. 

\newpage
\subsection*{A.2 Supplementary Figures}

\renewcommand{\thefigure}{S1}
\begin{figure}[h]
	\centering
	\includegraphics[scale=0.85]{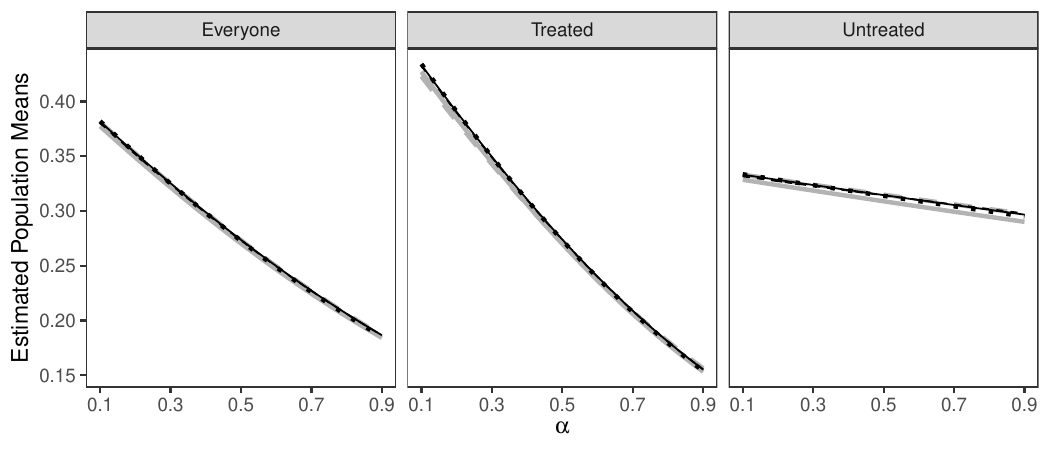}
	\caption{Estimates of the population mean estimands from the malaria bed net study dropping individual covariates. 
 Solid black lines represent the original scenario, solid gray lines represent dropping average age, dashed gray lines represent dropping average building materials, dashed black lines represent dropping the proportion of women, and dotted black lines represent dropping average urbanicity.}
	\label{fig: sensitivity drop individual covariates1}
\end{figure}

\renewcommand{\thefigure}{S2}
\begin{figure}[h]
    \centering
	\includegraphics[scale=0.85]{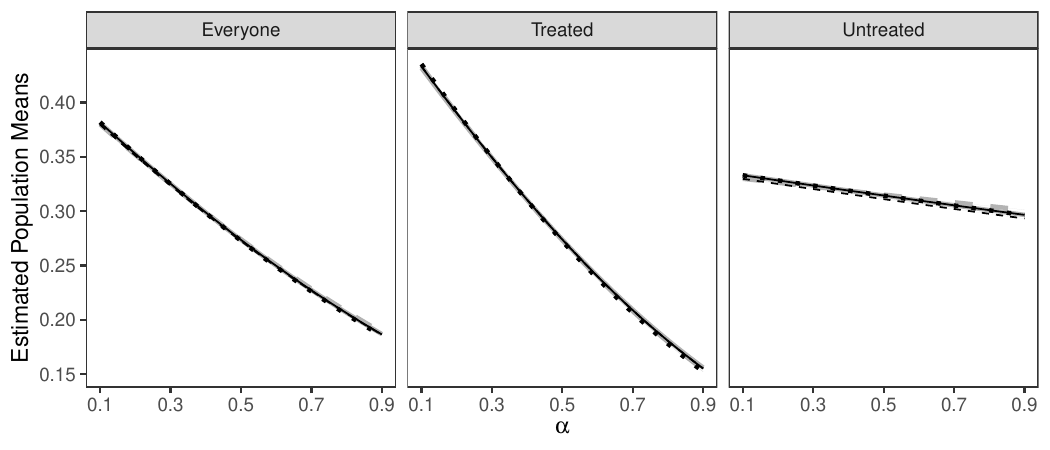}
	\caption{Estimates of the population mean estimands from the malaria bed net study dropping individual covariates. Solid black lines represent the original scenario, solid gray lines represent dropping average proportion of agricultural land cover, dashed gray lines represent dropping average altitude, dashed black lines represent dropping average total precipitation, and dotted black lines represent dropping average temperature.}
	\label{fig: sensitivity drop individual covariates2}
\end{figure}

\renewcommand{\thefigure}{S3}
\begin{figure}[h]
	\centering
	\includegraphics[scale=0.85]{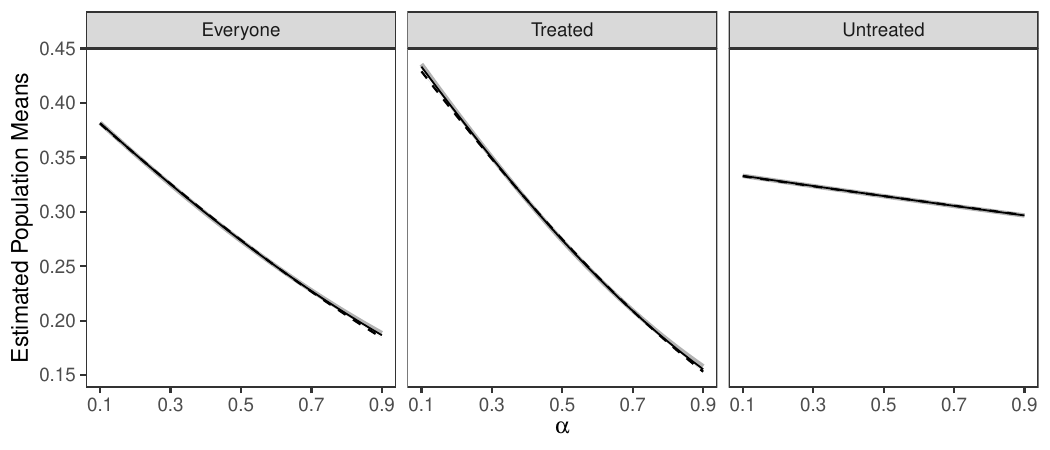}
	\caption{Estimates of the population mean estimands from the malaria bed net study changing the link function and adding an interaction term. The solid black lines represent the original scenario, the dashed black lines represent the model using probit link functions, and the solid gray lines represent the model with an interaction term.}
	\label{fig:sensitivity link and interaction}
\end{figure}

\end{document}